\documentclass[%
prc,%
twocolumn,%
reprint,%
superscriptaddress,%
floatfix,%
a4paper,%
twoside,
]{revtex4-2}

\usepackage[english]{babel}
\usepackage{graphicx}

\usepackage{amsmath}
\usepackage{amssymb}
\usepackage{amsfonts}
\usepackage{isotope}

\begin{document}
\title{Correlation studies of the $^{7}$He excited states}

\author{M.S.~Khirk}
\email{mskhirk@jinr.ru}
\affiliation{Flerov Laboratory of Nuclear Reactions, JINR,  141980 Dubna,
Russia}
\affiliation{M.V.~Lomonosov Moscow State University, Skobeltsyn Institute
of Nuclear Physics, 119991 Moscow, Russia}
\affiliation{MIREA - Russian Technological University, Moscow, 119454, Russia}

\author{L.V.~Grigorenko}
\affiliation{Flerov Laboratory of Nuclear Reactions, JINR,  141980 Dubna,
Russia}
\affiliation{National Research Nuclear University ``MEPhI'', 115409 Moscow,
Russia}
\affiliation{National Research Centre ``Kurchatov Institute'', Kurchatov sq.\ 1,
123182 Moscow, Russia}

\author{E.Yu.~Nikolskii}
\affiliation{National Research Centre ``Kurchatov Institute'', Kurchatov sq.\ 1,
123182 Moscow, Russia}
\affiliation{Flerov Laboratory of Nuclear Reactions, JINR,  141980 Dubna,
Russia}

\author{P.G.~Sharov}
\affiliation{Flerov Laboratory of Nuclear Reactions, JINR,  141980 Dubna,
Russia}
\affiliation{Institute of Physics, Silesian University in Opava, 74601 Opava,
Czech Republic}


\begin{abstract}
The unbound nucleus $^{7}$He was recently studied in the
$^2$H($^{6}$He,$^1$H)$^{7}$He
reaction at 29 $A\,$MeV beam energy in Ref.\ [M.\ S.\ Golovkov \textit{et al.},
Phys.\ Rev.\ C 109, L061602 (2024)]. The excitation spectrum of $^{7}$He  was
measured up to $E_T=8$ MeV ($E_T$ is energy above the $^{6}$He-$n$ threshold).
Angular distribution for the $^{6}$He-$n$ decay of the $^{7}$He  $3/2^-$ ground
state can be explained by a strong spin alignment induced by a reaction
mechanism. The correlation information for the  higher-lying $^{7}$He
excitations is available as backward-forward asymmetry for the $^{6}$He-$n$
decay in the $^{7}$He frame. The asymmetry function has an expressed energy
profile which may be explained by using quite restrictive assumptions about
structure of $^{7}$He excitations or/and peculiarities of the reaction
mechanism. In the analysis of [M.\ S.\ Golovkov \textit{et al.}, Phys.\ Rev.\ C
109, L061602 (2024)] the observation the $s_{1/2}$ state in $^{7}$He is declared
with $E_r \approx 2.0$ MeV. Our work is based on the
same $^{7}$He data. However, the data analysis was improved and also the data
interpretation is substantiated with the detailed PWBA reaction studies and
coupled-channel calculations of the $^{7}$He continuous spectrum. The idea of
the $s_{1/2}$ resonant state with $E_r \approx 2.0$ is rejected. In addition,
the position of the $1/2^-$ state in $^{7}$He is confined to the interval
$E_r=2.2-3.0$ MeV, with ``preferred'' value 2.6 MeV. There is indication on the
second $3/2^-$ state in the data with $E_r \sim 4.5$ MeV and with the lower
resonance energy limit $E_r \gtrsim 3.5$ MeV.  Importance and
prospects of more detailed correlation studies of $^{7}$He continuum are
discussed.
\end{abstract}


\maketitle

\section{Introduction}

Clear understanding of excitation spectra of the lightest nuclei is critical for
general understanding of nuclear structure and nucleon-nucleon interaction in
nuclei. The $^{7}$He is intriguing system in this sense. We know the $3/2^-$
ground state (g.s.) properties very well. The $5/2^-$ state at $E_T \sim 3$ MeV 
above the
$^{6}$He-$n$ threshold is clearly identified because of its unique dominating
decay pattern $^{4}$He+$3n$ \cite{Korsheninnikov:1999}. However, information
about other $^{7}$He excitations is nebulous.

The $^{7}$He has already been studied in the $(d,p)$ reaction in the works
\cite{Golovkov:2001,Wuosmaa:2005}. In comparison with the previous works a more
complex detector setup, which allows to detect \isotope[7]{He} decay products 
(and, thus, makes possible correlation studies) was used  in the experiment 
\cite{Bezbakh:2024,Golovkov:2024}. This work is based on the same $^{7}$He data 
as \cite{Bezbakh:2024,Golovkov:2024}, however, with improved analysis procedure 
and extensive theoretical discussion focused on correlation studies, see Secs.\ 
\ref{sec:experiment} and \ref{sec:corel}. The correlation data provide a 
distinctive rapid-varying behavior of the backward-forward asymmetry
function for population of the $^{7}$He continuum. Such a behavior of
this function may be explained by using quite restrictive assumptions as it is
strongly sensitive to the fine details of all the $3/2^-$, $1/2^-$, and $1/2^+$
configurations expected in the low-energy range of the $^{7}$He spectrum.

Generally, the $^{7}$He system has been studied many times and detailed reviews
of this work can be found in \cite{Renzi:2016} and \cite{Fortune:2018}. There is
the $3/2^-$ g.s.\ at $E_T=0.445$ MeV and some broad overlapping
structures are
typically observed above it. The evident candidate to be present in this
structures is the $1/2^-$ spin-orbit partner of the  $3/2^-$ ground state. The
review \cite{Fortune:2018} split the results in this field into two ``camps'':
(i) the works which support existence of the low-lying  $1/2^-$  excited state
with $E \lesssim 2$ MeV in $^{7}$He
\cite{Markenroth:2001,Meister:2002,Halderson:2004,Skaza:2006,Canton:2006,%
Ryezayeva:2006,Myo:2007,Arai:2009}, and (ii) the works which are in favor of the
$1/2^-$ with $E_T \gtrsim 2$ MeV
\cite{Poppelier:1985,Wolters:1990,Wurzer:1996,Navratil:1998,Bohlen:2001,%
Wuosmaa:2005,Mittig:2005,Aksyutina:2009,Pieper:2002,Baroni:2013}.
Such a dichotomy is a bit artificial, however it well demonstrates a strong
disagreement among different studies, both theoretical and experimental.
The $^{6}$He-$n$ correlation data obtained in this work indicate that the
$1/2^-$ state should be reasonably low-lying ($E_T \sim 2.2-3.0$ MeV) and should
be well populated (with population cross section, comparable with the cross
section for the $3/2^-$ g.s.).

The $3/2^-$ g.s.\ is well known to have non-single-particle structure with
important (even dominant) $^{6}$He$(2^+)$+$n$ wave function (WF) component.
Qualitatively this means that the excited states represented by the orthogonal
(to the g.s.) mixtures of the  $^{6}$He(g.s.)+$n$ and $^{6}$He$(2^+)$+$n$ WF
components may be situated reasonably low in excitation energy. There are
various theoretical calculations predicting the $3/2^-_2$ state of $^{7}$He at
$E_T \sim 3.5-6.0$ MeV \cite{Myo:2009,Baroni:2013,Rodkin:2021}. There is 
indication on the second $3/2^-_2$ state
in our $^{7}$He data with $E_r \sim 4.5$ MeV and with the lower resonance energy
limit $E_r \gtrsim 3.5$ MeV.

We also discuss possibility of the resonant $1/2^+$ state at about $E_T \sim
2-3$ MeV as predicted in \cite{Rodkin:2021} and observation of such a state in
$^{7}$He is declared in \cite{Golovkov:2024} with $E_r \approx 2.0$, $\Gamma 
\approx 2.0$ MeV. The possibility of the $1/2^+$ state
was for a long time discussed in $^{9}$He 
\cite{Chen:2001,Kalanee:2013,Golovkov:2007,Fortier:2007}. From qualitative point 
of view the principal possibility of such a state in $^{9}$He is supported by
evolution of the $s_{1/2}$ ``intruder'' orbital along the $N=7$ isotone: there
are ground states in $^{10}$Li, $^{11}$Be and low-lying excitations in $^{12}$B,
$^{13}$C build on the $s_{1/2}$ configuration. No such support for the low-lying
$1/2^+$ state of $^{7}$He can be found in the $N=5$ isotone: only in $^{9}$Be
there is a weak evidence for the $1/2^+$ state above the $^{8}$Be+$n$ threshold.
Theoretical works except the paper \cite{Rodkin:2021} all predict either no
low-lying $1/2^+$ state or repulsion in the $^{6}$He+$n$ channel 
\cite{Jaganathen:2017,Baroni:2013,Myo:2009,Myo:2020,Fossez:2018,Mazur:2022}. In
our work we demonstrate that the on-shell R-matrix parameterization used in the
work \cite{Golovkov:2024} to infer the $1/2^+$ resonance properties in $^{7}$He
is not applicable for the reaction studies (where the off-shell T-matrix should
be applied).

The situation when we do not understand such a basic structure characteristic as
spin-split between spin-orbit partners and the ground state multiplet structure
in such a ``simple'' light nuclide as $^{7}$He is quite unsatisfactory and it is
calling to a dedicated research. So, the aim of this work is to get a deeper
insight in these questions by using the correlation information obtained in the
$^2$H($^{6}$He,$^1$H)$^{7}$He reaction.

The system of units $\hbar=c=1$ is used in this work.


\section{Experimental setup and method}
\label{sec:experiment}

The experiment was carried out at the fragment separator ACCULINNA-2
\cite{Fomichev:2018} at U-400M heavy ion cyclotron, at Flerov Laboratory of
Nuclear Reactions (JINR, Dubna), using the method of double $p$-$^{6}$He and triple $p$-$^{6}$He-$n$ coincidences, developed in \cite{Bezbakh:2020a}.


\subsection{Experimental setup}

\begin{figure}
\centering
\includegraphics[width=1\linewidth]{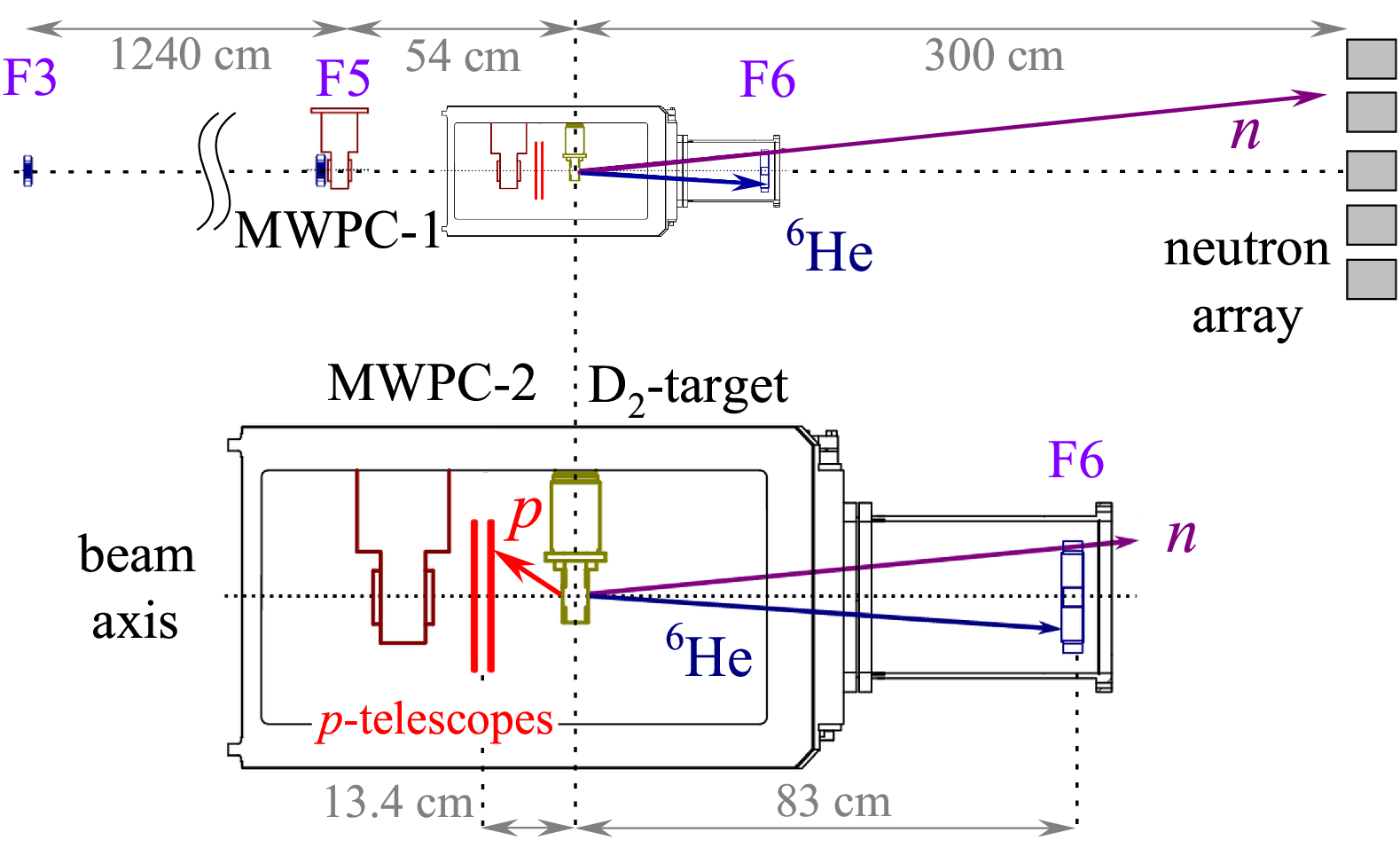}
\caption{
  Layout of the experimental setup. The ToF scintillator detectors are located
  at   F3, F5, and F6;  MWPC-1 and MWPC-2 --- beam tracking detectors. Arrows 
  show typical tracks of the $^2$H($^{6}$He,$^1$H)$^{7}$He reaction products.}
\label{fig:setup2}
\end{figure}

The consistant description of the experimental setup is presented
in \cite{Bezbakh:2024}.
Here we provide only a brief account of the main parts of the experimental
setup which we consider useful for the futher discussion.
Fig.~\ref{fig:setup2} shows schematic layout of the experimental setup with general
dimensions.

The 29~$A\,$MeV \isotope[6]He secondary beam
was produced with ACCULINNA-2 fragment separator by fragmentation of a
primary 33.4~$A\,$MeV \isotope[11]{B} beam  on beryllium target.
The standard ACCULINNA-2 scintillator time-of-flight (ToF) stations were
used for \isotope[6]{He} beam energy measurements for each event.
The multi-wire proportional chambers MWPC-1 and MWPC-2 provided
secondary beam track reconstruction.
A cryogenic cell filled with \(\isotope{D}_2\) gas with a tickness of
\(3.4(5)\times 10^{20}\) atom/cm\textsuperscript{2} was used as a
physical target.

The protons from the reaction were registered by four telescopes
(\(p\)-telescopes),
each consisting of two layers of double-sided silicon strip detectors.
The \(p\)-telescopes were  covered the
\(\sim\) 150\(^\circ\)--170\(^\circ\) range of laboratory angles.

\begin{figure}
\centering
\includegraphics{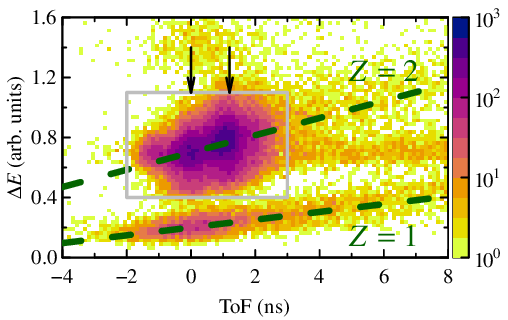}
\caption{
  The $\Delta E$-ToF ID plot for the F5--F6 ToF detectors.
  Dashed lines mark location of $Z=1$ and $Z=2$ heavy fragments.
  Arrows mark the locations of events connected with unreacted beam
  (left arrow, ToF $\equiv 0$) and \isotope[7]{He} g.s.\ (right arrow). The gray 
  rectangle indicates the data selection used in the analysis, see Fig.\ 
  \ref{fig:spectra}.}
\label{fig:6HeID}
\end{figure}

The ToF base F5--F6 was used for measurement of the heavy reaction fragment
(\isotope[4,6]{He}) longitudinal velocity.
Also the energy deposit in thin plastic at F6 was used for the heavy particle identification.
The two-dimensional time-amplitude spectrum of the F6 ToF-detector is shown in
Fig.~\ref{fig:6HeID}.
The resolution makes it possible to identify only \(Z\) of the heavy fragment.
However, this is sufficient to significantly suppress the
background level in $^{7}$He missing mass (MM) spectrum obtained in coincidence 
with the F6 detector.

The array of the 48 neutron detectors \cite{Bezbakh:2018} was used for
registration of neutrons originating from the \isotope[7]{He} decay.
Detection of a neutron from the \isotope[7]{He} decay allows
complete kinematics reconstruction for the $^2$H($^{6}$He,$^1$H)$^{7}$He
reaction. This gives additional opportunities for \isotope[7]{He}
structure studies discussed in the Sections \ref{subsec:exp-cm-method} and
\ref{subsec:exp-cor}.

\subsection{``Combined'' mass}
\label{subsec:exp-cm-method}


Important opportunity connected with neutron data is the complete-kinematics 
reconstruction of events. The registration of neutrons from the $^{7}$He decay 
provides information about its decay properties, in particular, the decay 
energy. The neutron-coincidence events formally contain redundant information 
about $^{7}$He decay energy and there exists an approach that allows to use this 
redundant information for improvement of the experimental resolution.

The following procedure is used. The velocity of the $^{7}$He center of mass 
(c.m.s.) is deduced from the momenta of the beam particle and the recoil proton. 
Using this velocity the momentum of neutron in $^{7}$He c.m.s.\ can be 
calculated. For the case of two body decay, the neutron momentum is sufficient 
for the $^{7}$He decay energy reconstruction.  This method of the decay energy 
reconstruction can be called ``combined mass method'', see \cite{Sharov:2017} 
for another application of such an approach. As far as the measurements of the 
neutrons' momenta are relatively precise and neutrons carry away the most of the 
decay energy, the combined mass spectrum reconstruction appears to be 
drastically more precise compared to the MM method. The Monte Carlo simulations 
of the experimental setup show that the energy resolution of the``combined mass 
method'' is about four times better than the MM energy resolution at energies 
around the \isotope[7]{He} ground state.


\subsection{Correlation measurements}
\label{subsec:exp-cor}


There are two important opportunities connected with registration of the
$p$-$n$-He coincidences:
(i) selection of events in the kinematical locus of the reaction can drastically
reduce the background level; (ii) due to the direct mechanism of the
$^2$H($^{6}$He,$^1$H)$^{7}$He reaction
the \isotope[7]{He} decay products momenta have specific correlations which are
sensitive to the \isotope[7]{He} structure.

As it is shown in Sec.\ \ref{sec:pwba}, in the frame of PWBA model the
reaction $^2$H($^{6}$He,$^1$H)$^{7}$He has two transferred momenta:
$\mathbf{q}_1$ (momentum transferred to spectator --- final state proton
momentum in the deuteron rest frame) and $\mathbf{q}_2$ (momentum transferred to
participant-target \isotope[7]{He} system --- deuteron momentum in the
\isotope[7]{He} center of mass frame). Within the PWBA model the  $\mathbf{q}_2$
vector provides the alignment direction around which the \isotope[7]{He} decay
products should be strongly correlated.

Ordinarily the center of mass angular distributions are used to deduce
spin-parities of the states populated in the direct reactions, see Refs.\
\cite{Golovkov:2001,Wuosmaa:2005} for the \isotope[7]{He} studied previously in
the $(d,p)$ reaction. However, in addition to that, the correlations in the
decay of particle-unstable \isotope[7]{He} can be studied to deduce this
information, see examples of such studies of $^{5}$H, $^{9}$He, and $^{10}$He
systems in Refs.\
\cite{Golovkov:2004,Golovkov:2005,Golovkov:2007,Sidorchuk:2010,Sidorchuk:2012}.

For the events with neutron coincidences (with complete kinematics) the
corresponding correlation functions can be directly extracted from data.
This analysis is discussed in the Sec.\ \ref{subsec:corel2}.

Because the experimental setup accepts only the events where the alignment
direction is approximately opposite to the beam direction, this effect should
show itself in the distribution of the heavy reaction fragment over the
longitudinal velocity. Analysis of this data is discussed in the Sec.\
\ref{sec:corel1}.


\section{Helium-7 excitation spectrum}


\begin{figure}
\centering
\includegraphics{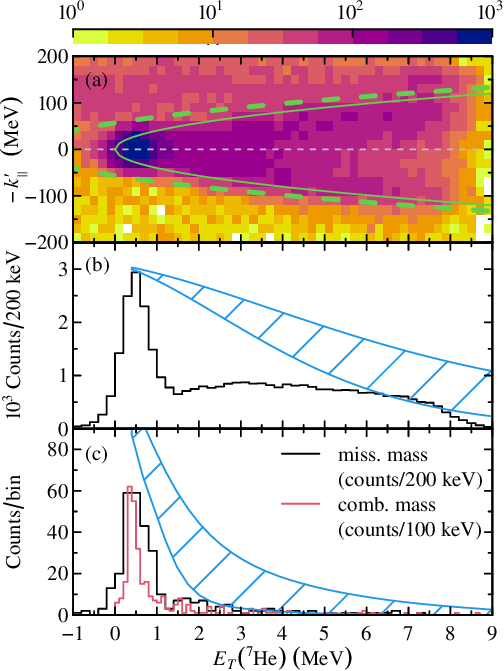}
\caption{(a) The heavy decay fragment longitudinal momentum in the 
\isotope[7]{He} center-of-mass system vs.\ \isotope[7]{He} missing mass. The 
solid curve shows the kinematical locus of the $^2$H($^{6}$He,$^1$H)$^{7}$He 
reaction. Dashed curve shows the kinematical cut used in the analysis according 
to the experimental resolution. (b) The \isotope[7]{He} MM spectrum. The hatched 
region shows the efficiency behavior according to MC simulations: the lower 
boundary was obtained assuming uncorrelated emission of $^{7}$He, and the upper 
boundary was obtained assuming maximum correlated parabolic (concave) 
distribution in the $Z||\mathbf{q}_2$ frame. (c) The $^{7}$He MM spectrum 
obtained in coincidence with neutrons and $^{7}$He combined mass spectrum. The 
hatched region shows efficiency behavior evaluated in the same way as in panel 
(b).}
\label{fig:spectra}
\end{figure}

With the experimental setup Fig.\ \ref{fig:setup2} a standard way to get the 
$^{7}$He spectrum is construction of its missing mass by using the information 
from the detection of a proton recoil. Figure \ref{fig:spectra} (a) shows the 
$^{7}$He MM augmented with information from the forward ToF detector --- 
registration of $^{6}$He and $^{4}$He heavy products from the $^{7}$He decay, 
see the gray rectangle in Fig.\ \ref{fig:6HeID}. The ToF data here is 
recalculated in terms of the heavy fragment longitudinal momentum 
$k'_{\parallel}$ in the $^{7}$He frame. It can be seen in this plot that the MM 
background conditions are quite good in the experiment, with the major 
background contribution coming from the random coincidences with the $^{6}$He 
beam located at about $k'_{\parallel} \sim 130-150$ MeV. The kinematical cut-off 
in this plot allows to drastically reduce the background in the MM $^{7}$He 
spectrum, see the result in Fig.\ \ref{fig:spectra} (b). It should be understood 
that $^{4}$He heavy products originating from the $^{7}$He $\rightarrow$ 
$^{6}\text{He}^*$+$n$ $\rightarrow$ $^{4}\text{He}$+$3n$ decay channel are 
confined inside the kinematical locus shown in Fig.\ \ref{fig:spectra} (a), and, 
thus the MM spectrum of Fig.\ \ref{fig:spectra} (b) includes this channel as 
well.

The energy resolution of the $^{7}$He MM spectrum in Fig.\ \ref{fig:spectra} (b) 
was obtained as $\sim 600$ keV FWHM in the energy range $0 < E_T< 2$ MeV and 
$700$ keV FWHM in the energy range $6 < E_T < 8$ MeV by the Monte-Carlo 
simulations. The results of the simulations are well confirmed by the observed 
width of the $^{7}$He ground state peak. The $^{7}$He MM spectrum has a sharp 
cut-off at about $ E_T \approx 8$ MeV connected with threshold for registration 
of the slow protons in the backward telescopes. The results of the setup 
efficiency simulations are shown in Fig.\ \ref{fig:spectra} (b) by the blue 
hatched regions. The simulations are shown with arbitrary scaling and normalazed 
on peak values at 0.4 MeV. The lower bounds of these regions correspond to 
assumption of isotropic $^{7}$He $\rightarrow$ $^{6}\text{He}^*$+$n$ decay. The 
upper bounds correspond to the most extreme correlated decay case when the 
products are strongly focused in the forward/backward direction in the frame 
aligned with transferred momentum, see this discussion further in Sec.\ 
\ref{subsec:corel2}. This strong focusing is described by pure $x^2$ term in 
Eq.\ (\ref{eq:parabol}). Using the efficiency corrections we can find that from 
$60-75 \, \%$ of the total cross section with $E_T \lesssim 6$ MeV are connected 
with population of excited states. In the first instance this may be seen as 
evidence for low spectroscopy of the $^{6}$He g.s.\ configuration in $^{7}$He 
g.s.\ structure.

A more advanced treatment of the $^{7}$He is available using the information 
from neutron coincidence data. It can be seen in Fig.\ \ref{fig:spectra} (c) 
that efficiency of the neutron wall is something around $3 \%$ on the $^{7}$He 
g.s.\ and it rapidly drops down at $E_T>1$ MeV rendering the spectrum 
practically nonexistent at $E_T \sim 2-3$ MeV. As it was discussed above in 
Section \ref{subsec:exp-cor} the combined mass spectrum reconstruction appears 
to be drastically more precise compared to missing mass providing the $\sim 150$ 
keV FWHM energy resolution for the $^{7}$He g.s., see the red histogram in Fig.\ 
\ref{fig:spectra} (c).

\begin{figure}
\centering
\includegraphics{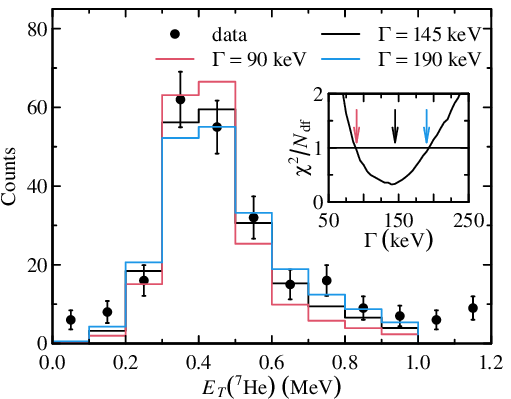}
\caption{Results of the MC simulations of the \isotope[7]{He} g.s.\ combined 
mass spectrum for different assumed widths (histograms) in comparation with the 
experimental spectrum (circles).   The insert shows \(\chi^2/N_{\text{df}}\) 
criteria as function of the \isotope[7]{He} g.s. width. The arrows in the insert 
correspond to the width values of the presented histograms.}
\label{fig:gs-width}
\end{figure}

The precise measurement of the $^{7}$He g.s. width may furnish additional information
about $^{7}$He structure. The widely used nuclear levels compilation~\cite{Tilley:2002} provides $\Gamma = 
0.15(2)$~MeV for the $^{7}$He g.s. In the further 
works~\cite{Beck:2007,Denby:2008,Aksyutina:2009,Cao:2012,Renzi:2016} the 
estimates of $\Gamma$ are provided in the range $0.12-0.19$~MeV. It should be 
noted that none of these works have made a crucial effort to determine this 
value precisely and disagreement among different experimental values is 
typically larger than their declared errors. The combined mass method gives a high 
enough resolution for the $^{7}$He g.s. width estimation. Such kind of estimates have been given in \cite{Bezbakh:2024}.
In this paper we present the results of more enhanced analysis of the $^{7}$He g.s. parameters.
The $^{7}$He combined mass spectrum 
is shown in Fig.~\ref{fig:gs-width} on a larger scale together with MC 
simulations for different $^{7}$He g.s.\ widths values. The analysis of the $\chi^2$ 
analysis, see the inset in Fig.\ \ref{fig:gs-width}, gives estimates of the resonance energy 
$E_r = 0.41(2)$ and width $\Gamma = 0.14(5)$ MeV. The errors (confidence intervals)
we define by a criterion $\chi^2/N_{\text{df}} = 1$.  One can see that the new results is quite close to the results
of \cite{Bezbakh:2024} (\(E_r = 0.38(2)\) and \(\Gamma = 0.11(3)\) MeV). However the new \(\Gamma\) estimate have greater value and it confidence interval is significantly wider, therefore it's consistent with the results of the majority of previous works. We consider that obtained precision for \(\Gamma\) value is mainly determined not by measurement method, but by the available statistics. According to our estimates, in a prospective experiment in analogous technique
(with reasonable statistics) the precision \(\Delta\Gamma \sim\) 10--15 keV may be aimed.

\section{PWBA model for the $^2$H($^6$He,$^1$H)$^7$He reaction}
\label{sec:pwba}

\begin{figure}
  \centering
\includegraphics[width=\linewidth]{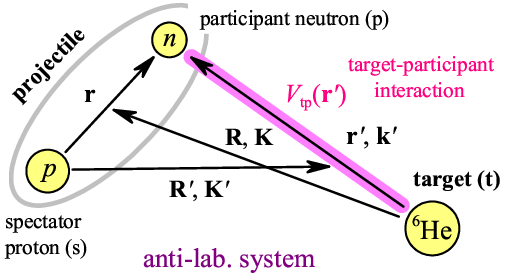}
\caption{Prior and post Jacobi coordinate settings for the PWBA model in the
(historically motivated) anti-lab system. Only one final state interaction is
taken into account in T-matrix in this approximation.}
\label{fig:coord-dwba}
\end{figure}

Important qualitative features of the $^2$H($^6$He,$^1$H)$^7$He
reaction can be understood based on the plane-wave Born approximation (PWBA)
model, which well reflects the ``quasifree scattering'' aspect of the process.
Within the historical terminology for this class of reactions we consider the
system of three particles, which, in the anti-lab system, are ``target'' $M_t$
($^{6}$He), and composite ``beam'' $M_b$ ($^{2}$H), consisting of ``spectator''
$M_s$ (proton), ``participant'' $M_p$ (neutron), see Fig.\ \ref{fig:coord-dwba}. 
The experiment is performed in the inverse kinematics and we keep
``beam'' and ``target'' notation with quotation marks for this Section to avoid
confusion here. Completely analytical treatment of the model is possible under
two essential physical assumptions:

\begin{itemize}

\item Plane wave motion is assumed in the initial  $\{\mathbf{R},
\mathbf{K} \}$ and final state coordinates $\{\mathbf{R}', \mathbf{K}' \}$.

\item Only one interaction is taken into account in the T-matrix --- the
potential $V_{\text{tp}}$ between ``target'' and ``participant'' fragment.

\end{itemize}

There is some confusion in terminology, but it looks that the most widespread
reading is that quasifree scattering (QFS) is further simplification of PWBA,
when the off-shell T-matrix is replaced with the experimental or
phenomenological cross section (on-shell T-matrix), sometimes with simple
extrapolation in the off-shell region.

Since 1950s \cite{Sitenko:1959,Shapiro:1961} this approach was elaborated on
numerous occasions \cite[and Refs.\ therein]{Frobrich:1996}.
It was especially popular for reactions with nucleon transfer from deuteron,
having the low binging energy and huge geometric extent, which make the above
physical approximations easy to justify.
Nevertheless, we found it necessary to provide a detailed discussion of the
model in this work for several reasons: (i) it is difficult to find in the
literature consistent discussion of the method in the aspect concerning
correlations in the decay of unstable product; (ii) the approach has never been
discussed in sufficient details in our previous experimental works
\cite{Golovkov:2004,Golovkov:2005,Golovkov:2007,Sidorchuk:2010,Sidorchuk:2012},
(iii) the rather detailed discussion of correlations in this work would be hard
to follow otherwise.

Figure~\ref{fig:coord-dwba} shows the  Jacobi coordinates for prior and post
forms of the wave function. These sets have a simple relations
\begin{align*}
  \mathbf{r}' &= \mathbf{R} + \alpha \mathbf{r} \;, \quad \alpha=
\frac{M_s}{M_s+M_p} ,\quad   \beta = \frac{M_t}{M_t+M_p}\;,
 \nonumber \\
\mathbf{R}' &= - \beta \mathbf{R}+ \gamma \mathbf{r} \;, \quad \gamma =
\frac{M_p (M_t+M_p+M_s)}{(M_s+M_p)({M_t+M_p})}\;.
 \nonumber
\end{align*}
The corresponding Jacoby momenta in this notation have the following meaning:
\(\mathbf K\) is the ``beam'' momentum in the reaction c.m.s.\
(the deuteron momentum for the case of the current experiment);
\(\mathbf {K}'\) --- momentum of the ``target''-participant subsystem
in the reaction c.m.s. (\isotope[7]{He} momentum);
\(\mathbf {k}'\) --- momentum of the participant (neutron) in the
``target''-participant subsystem c.m.s. With the plane-wave in- and out- WFs
\[
\Psi_{tb}   = \chi_{\mu_b} e^{i \mathbf{K} \mathbf{R}}  , \qquad
\Psi_{s(tp)}   =  \chi_{\mu_s} e^{i \mathbf{K}' \mathbf{R'}} ,
\]
the T-matrix has explicitly factorized form (the $^{6}$He ``target'' particle is
spinless)
\begin{multline}
T_{\mu'_p,\mu'_s,\mu_b}(\mathbf{K}',\mathbf{k}',\mathbf{K}) = \int d^3 r \, d^3
R \,\,
\Psi_{tp}^{\dagger}
(\mathbf{k}',\mathbf{r}')  \\
\times \, \Psi_{s(tp)}^{\dagger}(\mathbf{K}',\mathbf{R}') \, V_{tp}(\mathbf{r}')
\Psi_{sp} (\mathbf{r}) \, \Psi_{tb}(\mathbf{K},\mathbf{R}) = \\
= \sum_{\mu_p} T^{\dagger}_{\mu'_p,\mu_p}(\mathbf{k}',\mathbf{q}_2) \;
\Phi_{\mu'_s,\mu_p,\mu_b} ( \mathbf{q}_1 ) \;,
\label{eq:t-qfs-1}
\end{multline}
where the initial state effects are realized via formfactor $\Phi$, while
the ``target''-participant interaction properties are reflected in the off-shell
T-matrix $T(\mathbf{k}',\mathbf{q}_2)$ of the ``quasifree subsystem''.

\begin{figure}
\centering
\includegraphics[width=\linewidth]{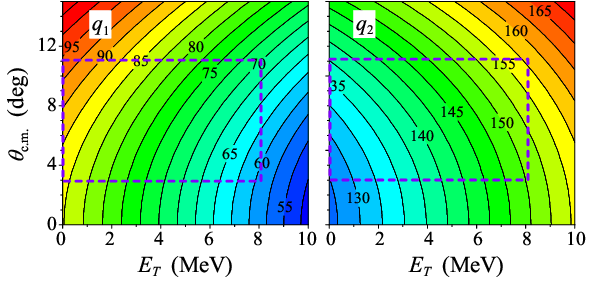}
\caption{The transferred momenta $\mathbf{q}_1$ and $\mathbf{q}_2$ as a function
of $E_T$ and $\theta_{\text{c.m.}}$. The violet rectangles show the kinematical
range accessible in the experiment.}
\label{fig:q1-q2-et-tetcm}
\end{figure}

It should be emphasized that there are \emph{two transferred momenta} in the
model: $\mathbf{q}_1$ (momentum transfer to spectator in the projectile rest
frame) and $\mathbf{q}_2$ (momentum transfer to the quasifree channel).
These momenta are defined by relation between initial and final state Jacobi
coordinates:
\begin{equation}
\mathbf{q}_1 = \alpha \mathbf{K} + \mathbf{K}' \quad , \qquad
\mathbf{q}_2= \mathbf{K}+\beta \mathbf{K}' \;.
\label{eq:q1-q2}
\end{equation}
The behavior of the transferred momenta within the kinematical range of interest
for the experimental conditions is illustrated in Fig.\
\ref{fig:q1-q2-et-tetcm}.

The transfer formfactor (the simple $s$-wave deuteron WF $\Psi_{sp}$ is assumed)
and the elastic scattering $T$-matrix are defined in a standard way
\begin{align}
\Phi_{\mu'_s \mu_p,\mu_b}(\mathbf{q}) = C^{1 \mu_b}_{1/2 \mu'_s 1/2 \mu_p} \int
d^3 r \; \chi^{\dagger}_{\mu'_s} \chi^{\dagger}_{\mu_p} e^{-i \mathbf{qr}} \,
\Psi_{sp}(\mathbf{r}) \;, \nonumber \\
T_{\mu'_p,\mu_p}(\mathbf{k}',\mathbf{k}) =\int d^3 r \;
\chi^{\dagger}_{\mu'_p}e^{-i \mathbf{k'r} } \,
V_{tp}(\mathbf{r})\, \psi_{tp} (\mathbf{k},\mathbf{r}) \;.\quad
\end{align}
%

The triple-differential cross section is
\begin{multline}
\frac{d \sigma}{dE_T \, d \Omega_{K'}} = \frac{M_{bt}\,M_{s(tp)}}{(2\pi)^2} \;
\frac{K'}{K} \; \frac{2M_{tp} k'}{\pi} \;  \\
\times  \Bigl[ (4\pi)^2 \; |\phi_{0}(q_1)|^2  \sum_{jl} \frac{2j+1}{2s_p+1}
\, |T_{jl}(q_2,k')|^2 \Bigr]\,,
\label{eq:cross-sec-2dmmm}
\end{multline}
and five-fold-differential cross section suitable for $^{7}$He correlation
studies can be presented in the density-matrix formulation, which is specially
convenient for the further phenomenological studies
\begin{multline}
\frac{d \sigma}{dE_T \, d \Omega_{K'} \, d \Omega_{k'}}  =
\frac{ M_{bt} M_{s(tp)} }{(2\pi)^2} \; \frac{K'}{K} \; \frac{2M_{tp}k'}{\pi} \\
\times  \; \sum_{j'l'm'_j} \sum_{j l m_j} \; \rho^{j'l'm'_j}_{j l m_j} \;
\sum_{\mu'_p} A^{\dagger}_{j'l'm'_j \mu'_p} \; A_{j l m_j \mu'_p} \,.
\label{eq:cs-triple-dif}
\end{multline}
The density matrix depends on the excitation energy
\[
E_T=\frac{k'^2}{2M_{tp}} \, ,
\]
and the center of mass reaction angle $\theta_{\text{c.m.}}$, associated with
the momentum $\mathbf{K}'$, while the transition amplitudes depend on
$\theta_{\text{c.m.}}$ and $\mathbf{k}'$
\begin{align}
\rho(E_T,\theta_{\text{c.m.}})^{j'l'm'_j}_{jlm_j}= (4\pi)^2 \; |\phi_{0}(q_1)|^2
\; M^{j'l'm'_j}_{jlm_j} \,, \quad \nonumber \\
A_{j l m_j \mu'_p}(\theta_{\text{c.m.}},\mathbf{k}') = \frac{\hat{j}}{\hat{s}_p}
\, T_{jl}(q_2,k')
\sum_{ m } C^{jm_j}_{lm s_p \mu'_p}  \; Y_{lm}(\hat{k}') \,.\,\quad
\label{eq:rho-a-def}
\end{align}
The partial components of the elastic $T$-matrix and transfer formfactor for the
simple $s$-wave deuteron WF with radial component $\psi_{0}(r)$ are
defined as
\begin{align}
T_{jl}(k',k) =\frac{1}{k'k} \int  dr \; F_l (k'r) \,
V_{tp}(r)\, \psi_{jl} (kr) \;,
\label{eq:t-def} \\
\phi_{0}(q) = \frac{1}{q} \int d r \; F_0 (qr)  \, \psi_{0}(r) \;.
\label{eq:phi-def}
\end{align}
The matrix $M$ is pure angular density matrix
\begin{equation}
M^{j'l'm'_j}_{jlm_j}\!=\frac{4 \pi}{\hat{j}'\hat{j}}\!
\sum_{m'm \mu_p}\! C^{j'm'_j}_{l'm's_p \mu_p}
C^{jm_j}_{lm s_p \mu_p}  Y^*_{l'm'}(\hat{q}_2)  Y_{lm}(\hat{q}_2),
\label{eq:m-matr}
\end{equation}
with symmetries
\begin{equation}
\langle j'l'm| \rho | jlm \rangle = (-)^{l+l'-j-j'+1} \; \langle j'l'-m| \rho |
jl-m \rangle \;,
\label{eq:eq:den-matr-symmetr}
\end{equation}
It is normalized to give unity for each $\{j,l\}$ state
\[
\textstyle \sum_{m_j}  M^{jlm_j}_{j l m_j} \equiv 1 \,.
\]
The maximum spin alignment in the ``quasifree'' $^{7}$He channel is realized in
the system, where $Z||\mathbf{q}_2$. In this frame the density matrix gets the
most sparse and simple form. For the $\{s_{1/2},p_{1/2},p_{3/2} \}$ state vector
one gets
\[
  M^{j'l'm'_j}_{jlm_j} =
  \begin{pmatrix}
    1/2  & 0   & -1/2 & 0    &  0  & 1/2  & 0   &  0  \\
    0    & 1/2 &    0 &  1/2 &  0  & 0    & 1/2 &  0  \\
    -1/2 & 0   &  1/2 &    0 &  0  & -1/2 & 0   &  0  \\
    0    & 1/2 &    0 &  1/2 &  0  & 0    & 1/2 &  0  \\
    0    & 0   &    0 &  0   &  0  & 0    & 0   &  0  \\
    1/2  & 0   & -1/2 &    0 &  0  & 1/2  &  0  &  0  \\
    0    & 1/2 &    0 &  1/2 &  0  & 0    & 1/2 &  0  \\
    0    &  0  &    0 &  0   &  0  & 0    & 0   &  0
  \end{pmatrix} \,.
  %
\]

The scattering WF $\psi_{jl} (kr)$ diagonalizing the elastic $S$-matrix, is
normalized by the asymptotic condition
\[
\psi_{jl} (kr) \rightarrow e^{i \delta_{jl}}\; \sin(kr-l\pi/2+\delta_{jl})\;.
\]
For that reason the PWBA transition amplitude can be represented as
\begin{equation}
\frac{\hat{j}}{\hat{s}_p} \; T_{jl}(q_2,k') = a_{jl}(q_2,k') \; e^{i
\delta_{jl}(E_T)} \;,
\label{eq:amp-repres}
\end{equation}
where the $a_{jl}$ are real-valued functions. For reactions well suited for the
quasifree approximation, we may expect, although can not be completely
confident, that the real transition amplitudes and phases are reasonably close
to the PWBA values. The population probabilities for different states can be
defined as
\begin{equation}
W_{jl}(q_2,k')=S_{jl}\; \frac{2M_{tp}k'}{\pi} \; a_{jl}^2(q_2,k') \;,
\label{eq:wjl}
\end{equation}
where phenomenological corrections, if necessary, are included via spectroscopic
factors $S_{jl}$.

The correlations pattern for the $\{s_{1/2},p_{1/2},p_{3/2} \}$ state set in the
 frame where $\mathbf{q}_2$ directed along \(Z\) axis can be obtained as (case 
 of a complete equatorial alignment)
\begin{multline}
\frac{d \sigma}{dx} \sim  \frac{1}{2} \, a^2_{s_{1/2}} + \frac{1}{2} \,
a^2_{p_{1/2}} + \frac{1}{4} \, a^2_{p_{3/2}}\,(1+3x^2) \\
+ \, \frac{1}{\sqrt{2}} \, a_{p_{1/2}} a_{p_{3/2}}\, (3x^2-1)
\,\cos(\delta_{p_{1/2}}^{p_{3/2}})  \\
+ \, a_{s_{1/2}} \,x\, \left[a_{p_{1/2}} \cos(\delta_{s_{1/2}}^{p_{1/2}}) +
\sqrt{2} a_{p_{3/2}} \cos(\delta_{s_{1/2}}^{p_{3/2}}) \right]  \,,
\label{eq:corel-x-all}
\end{multline}
where the $x$ variable and the relative phases are
\begin{equation}
x = \cos(\theta_{k'})\equiv\cos(\widehat{\mathbf{k}',\mathbf{q}_2})\;,\quad
\delta_{j_2l_2}^{j_1l_1}=\delta_{j_1l_1}-\delta_{j_2l_2} \,.
\label{eq:x-delta}
\end{equation}
%

\begin{figure}
\centering
\includegraphics[width=\linewidth]{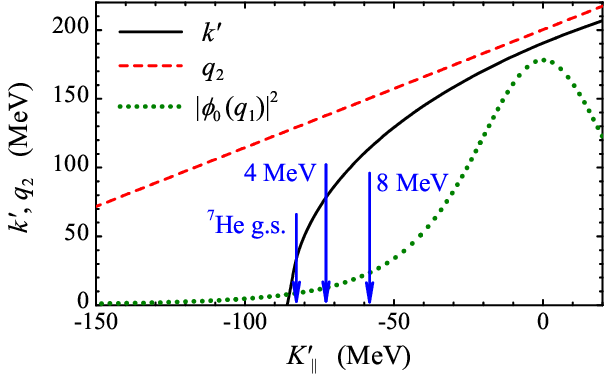}
\caption{The values of $\mathbf{k}'$ and $\mathbf{q}_2$ as a function of lab
momentum of the recoil proton $K'_{\parallel}$ at $K'_{\perp} \equiv 0$, which
also corresponds to the reaction lab angle $\theta_{\text{c.m.}}=0$ and relation
$q_1=K'_{\parallel}$. The $^{7}$He threshold energies $E_T=\{0.45,4,8\}$ MeV are
indicated by the vertical blue arrows. The deuteron formfactor
$|\phi_{0}(q_1)|^2$ (green dotted curve) is shown with arbitrary scaling.}
\label{fig:three-momenta}
\end{figure}

There are two important features clarified within the PWBA model at that point:

\begin{itemize}

\item The $(d,p)$ reaction for our kinematical conditions are not 
``comfortable'' for interpretation in terms of ``quasifree scattering''. The
values of $\mathbf{k}'$ and transferred momentum $\mathbf{q}_2$ entering the
$T(q_2,k')$ are shown in Fig.\ \ref{fig:three-momenta}. The off-shell
effect is large in this reaction and this effect is strongly varying across the
$^{7}$He excitation energy range $E_T \lesssim 8$ MeV accessible in the
experiment. According to Fig.\ \ref{fig:three-momenta} one may expect that
quasifree scattering approximation may become safe (small off-shell corrections)
at $E_T \sim 20-30$ MeV.

\item  In the energy range of interest, the deuteron formfactor
$|\phi_{0}(q_1)|^2$ value vary by factor 2--3. Thus, the population of the
$^{7}$He excitation spectrum does not provide the direct information on
spectroscopy of $^{7}$He excited states: within PWBA model this information for
broad excited state of \isotope[7]{He} is expected to be strongly effected by
the reaction mechanism.

\end{itemize}


\section{Models for the $^7$He continuum}
\label{sec:7he-models}


The easiest approximation for the $^{7}$He continuum states are provided by the
single-channel potential model. For simplicity we used the square-well
potential, which is sufficient for qualitative considerations and allows to make
most of calculations analytically.

\begin{figure}
\centering
\includegraphics[width=0.45\textwidth]{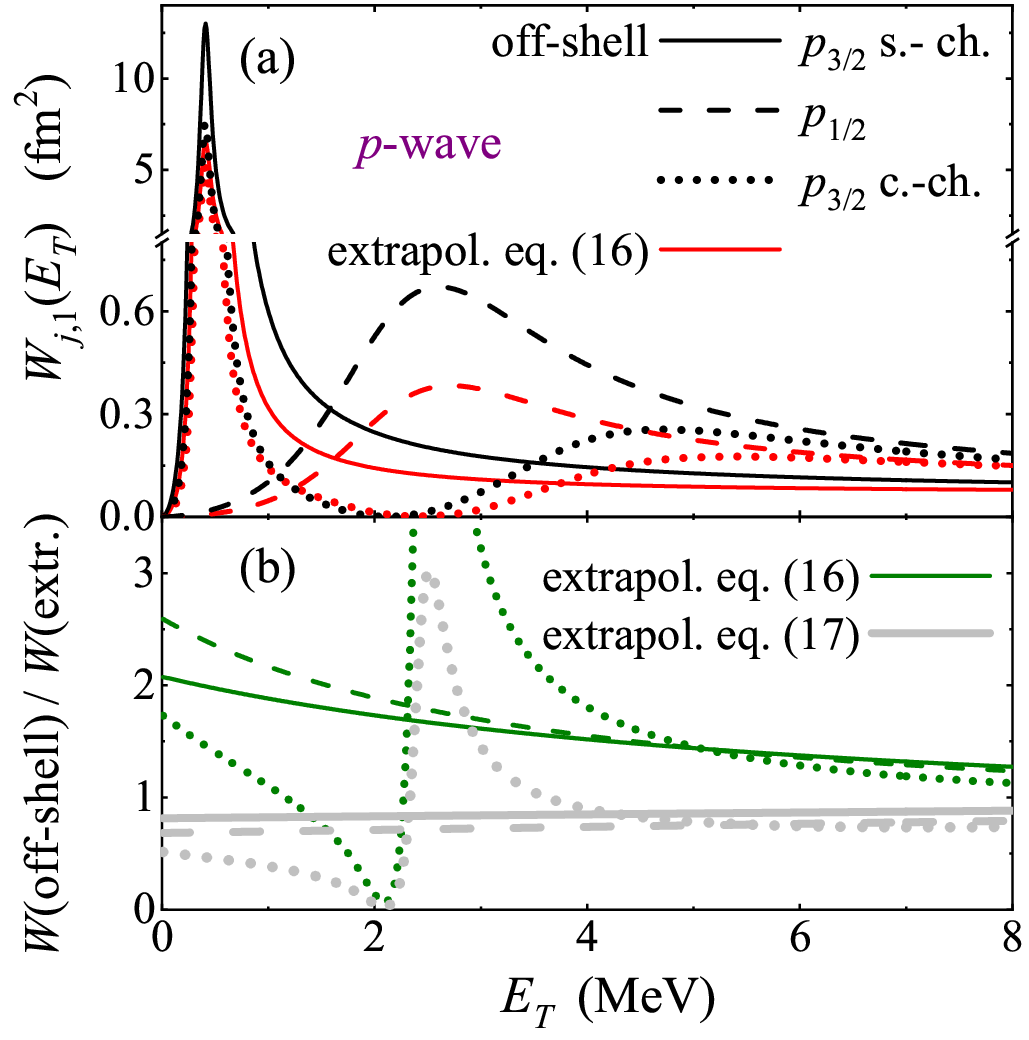}
\caption{(a) The $l=1$ population probabilities Eq.\ (\ref{eq:wjl}) calculated
off-shell at $\theta_{\text{c.m.}}=8^{\circ}$, where most of the data are
located, and extrapolated from the on-shell
value by Eq.\ (\ref{eq:w-off1}).
(b) Ratios of the directly calculated off-shell probabilities to the two variants of the extrapolated
from the on-shell values, by Eq.\ (\ref{eq:w-off1}) and by Eq.\ (\ref{eq:w-off2}).
Calculations in the single-channel potential model for the $p_{3/2}$
and $p_{1/2}$ resonances and the coupled-channel model for the $p_{3/2}$ resonance are shown 
in the panels by solid, dashed and dotted lines, respectively.}
\label{fig:p-wave-off}
\end{figure}


\subsection{The $p_{3/2}$ ground state}


The $^{7}$He ground state width is around 140 keV.
With potential of reasonable radius $r_0=3$ fm one get width of $\sim 250$ keV.
To play with the width values in potential model we can vary the potential
width. This is not quite consistent with physics of the case, where reduction on
the $^{7}$He g.s.\ is connected with not single particle nature of this state
(strong mixing with $^{6}\text{He}(2^+)$+$n$ configuration). The $p_{3/2}$
continuum profile calculated with $r_0=1.86$ fm and giving
$\Gamma=140$ keV are shown in Fig.\ \ref{fig:p-wave-off} (a). In this
model the high-energy ``potential tail'' of the $p_{3/2}$ resonance is scaled as
$\sim r_0$. This behavior is important for understanding of the $p_{1/2}$ state
properties (discussed below) and correlations (see Section \ref{sec:p32-mod}).

Within the QFS approximation to PWBA \cite{Bezbakh:2024,Golovkov:2024} instead
of the off-shell T-matrix the phenomenological on-shell cross section is used,
which is extrapolated off-shell in some reasonable way. The on-shell population
probability Eq.\ (\ref{eq:wjl}) is related to the elastic cross section
$\sigma^{\text{(el)}}$ as
\begin{equation}
W_{jl}(k',k') = \frac{\sin^2 \delta_{jl}}{2 \pi M_{tp} k'}
= \frac{k' \sigma_{jl}^{\text{(el)}}(E_T)}{8 \pi^2 M_{tp}} \,.
\label{eq:w-sig-el}
\end{equation}
From the T-matrix definition (\ref{eq:t-def}) a reasonable off-shell
extrapolation is
\begin{eqnarray}
W_{jl}(q_2>k',k')& =&  \frac{k' P_l(q_2)}{q_2 P_l(k')} \, W_{jl}(k',k') 
\label{eq:w-off1}\\
& \approx & \left(\frac{q_2}{k'}\right)^{2l}  W_{jl}(k',k')\, ,
\label{eq:w-off2}
\end{eqnarray}
see, e.g., \cite{Fuchs:1972,Frobrich:1996}. It is evidently precise in the low
$q_2$ limit, which is not a good approximation for the experimental conditions,
see Fig.\ \ref{fig:three-momenta}. Nevertheless, it can be found in Fig.\
\ref{fig:p-wave-off} (b) that the off-shell extrapolation is quite precise in
the case of $p$-wave states, represented by potential scattering.

\begin{figure}
\centering
\includegraphics[width=0.45\textwidth]{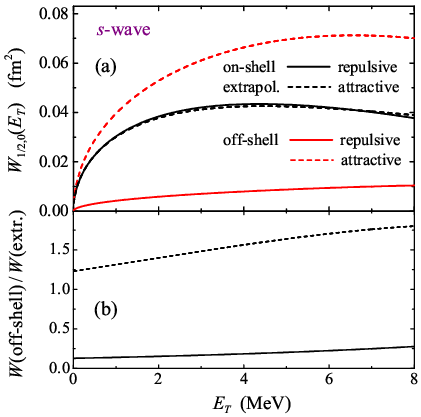}
\caption{(a) The $s$-wave population probabilities Eq.\ (\ref{eq:wjl})
calculated
off-shell at $\theta_{\text{c.m.}}=8^{\circ}$, and extrapolated from the
on-shell value by Eq.\ (\ref{eq:w-off1}). (b) Ratios of the directly calculated
off-shell probabilities to the extrapolated on-shell values.}
\label{fig:s-wave-off}
\end{figure}


\subsection{The $s_{1/2}$ state}


It is typically expected that interaction in this partial wave is quite 
featureless repulsion. The $s_{1/2}$ continuum profiles calculated with 
repulsive potential and deep attractive potential (with ``forbidden state'') are 
shown in Fig.\ \ref{fig:s-wave-off}. The phase shifts for these cases are 
fine-tuned to be very close to each other and to the phase shifts obtained in 
sophisticated continuum shell-model calculations \cite{Baroni:2013,Mazur:2022}. 
One may see in Fig.\ \ref{fig:s-wave-off} that the off-shell T-matrices are (i) 
strongly (factor $5-8$) different depending on specific dynamics and (ii) they 
are evidently not ``off-shell extrapolatable'' by simple expressions like Eq.\ 
(\ref{eq:w-off1}). This is demonstration that treatment of the $s$-wave 
populations in reactions should be considered with more caution. This fact is 
also important for discussion of possible interpretation of correlations in 
terms of $s$-wave resonance contributions, see Sec.\ \ref{sec:s12-mod}.

For attractive potential the PWBA $s$-wave population is on the upper limit of
what is admissible from experimental data. So, phenomenologically this
population can not be increased, but can be reduced (e.g.\ repulsive potential).
From correlations point of view if we reduce the $s$-wave population of Fig.\
\ref{fig:s-wave-off} 50-fold, it will be still sufficient to provide
experimentally observed backward-forward asymmetry, see Sec.\ \ref{sec:corel1}.

\begin{figure}
\centering
\includegraphics[width=0.46\textwidth]{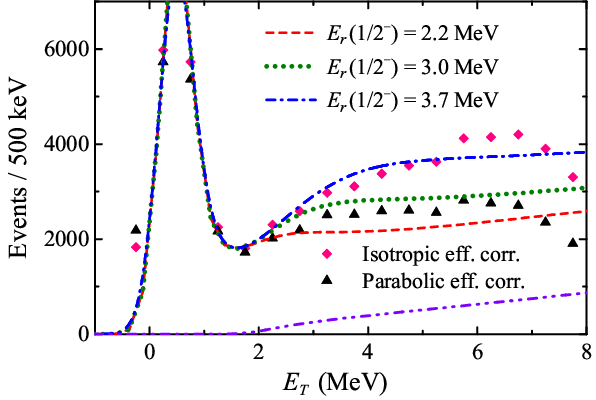}
\caption{The $^{7}$He state profiles provided by potential model for different
positions of the $p_{1/2}$ resonance (attractive $s$-wave potential) are
compared with the data, see also Fig.\ \ref{fig:ass-poten} (b). The data are
efficiency corrected by MC simulations in two ways: assuming uncorrelated
emission of $^{7}$He and maximally correlated parabolic (concave) distribution
in the $Z||\mathbf{q}_2$ frame (see also Fig.\ \ref{fig:spectra}). The possible
contribution of the decay channels $^{7}$He $\rightarrow$ $^{6}\text{He}^*+n$,
connected with disintegration of $^{6}$He $\rightarrow$ $\alpha$+$n$+$n$ are
extrapolated by using the data from \cite{Wuosmaa:2005}. }
\label{fig:comp-spec-all}
\end{figure}


\subsection{The $p_{1/2}$ state}


Quite reliable evidence for this state may be found in the
data as a deep in the $^{7}$He spectrum at $E_T=1.5-1.8$ MeV and corresponding
kink in the spectrum at $E_T\sim 3$ MeV. However, the $1/2^-$ peak is sitting on 
a strong ``background'' of the $3/2^-$ and $1/2^+$ contributions, which are in
general case comparable and not completely fixed by theoretical considerations.
One of the fits to the data (attractive $s$-wave potential) is shown in Fig.\
\ref{fig:comp-spec-all}. The spectroscopic factors for the $1/2^-$ state are
deduced by requesting a good fit quality in the energy range $E_T\sim 1-3$ MeV 
are listed in the Table \ref{tab:p12-populations}. The $1/2^-$ is expected to be 
a single-particle state with spectroscopic factor close to unity. Requesting the
spectroscopic factor to be in the range $S_{p_{1/2}} \sim 0.7-1.3$ we get limits
on possible  $1/2^-$ spectrum properties. The $1/2^-$ resonant energy positions
outside the $E_r=2.2-3.0$ MeV interval are not acceptable. Considering the
results of Sec.\ \ref{sec:p32-mod} as preferable explanation for the observed
correlations, $E_r \sim 2.6$ MeV seem to be a preferable value.

\begin{table}[b]
\caption{Spectroscopic factors $S_{p_{1/2}}$ for the $1/2^-$ state of $^{7}$He
obtained by fitting the experimental MM spectrum. The spectroscopic factor for
the $3/2^-$ g.s.\ is taken as $S_{p_{3/2}}=0.5$. Abbreviation FC stand for
``Fermi cut'' of the $3/2^-$ population probability at $E_T \sim 1.2-1.4$ MeV,
see also Section \ref{sec:p32-mod}.}
\begin{ruledtabular}
\begin{tabular}[c]{ccccc}
$E_r(1/2^-)$  & Attr.\ $s_{1/2}$ &  Rep.\ $s_{1/2}$  & Attr.\ $s_{1/2}$+FC &
Rep.\ $s_{1/2}$+FC  \\
\hline
2.2  & 0.4 & 0.6 & 0.9 & 1.1  \\
2.6  & 0.6 & 0.8 & 1.2 & 1.4  \\
3.0  & 0.9 & 1.2 & 1.4 & 1.8  \\
3.7  & 1.8 & 2.0 & 2.8 & 3.2  \\
\end{tabular}
\end{ruledtabular}
\label{tab:p12-populations}
\end{table}


\subsection{The coupled-channel $p_{3/2}$ state}


Important aspects of the continuum dynamics beyond the single-channel potential
approximation can be treated within the coupled-channel formalism. For our
purposes we construct the couple-channel model is quite a schematic way. The
coupled-channel Shr\"odinger equations look like
\[
  \begin{cases}
    \left( \hat{T}-E_T+\hat{V}_{11}\right)\Psi _{\isotope{He}\text{-}n}+
    \hat{V}_{12} \Psi _{\isotope{He}^*\text{-}n} = 0 \\
    \left( \hat{T}-(E_T-\Delta E_T)+\hat{V}_{22}\right)
    \Psi _{\isotope{He}^*\text{-}n}+
    \hat{V}_{12} \Psi _{\isotope{He}\text{-}n}=0  \, .
  \end{cases}
\]
The energy $E_T$ is considered from the $^{6}$He(g.s.)-$n$ threshold and $\Delta
E_T = 1.8$ MeV is the threshold difference with the $^{6}$He($2^+$)-$n$ channel.
The $^{6}$He($2^+$)-$n$ threshold is $\Delta E_T = 1.8$ MeV higher, and
it should be understood
that this notion is a certain approximation as the \isotope[6]{He}($2^+$) state
is particle unstable. The actual first threshold in the \isotope[7]{He} system
is the three-body $^{4}$He+$3n$ threshold at $E=0.97$ MeV. The dynamical issue
which saves the situation is that below the $^6$He($2^+$)-$n$ threshold the
decay mechanism for $^{7}$He states is so-called ``true'' $3p$ emission, which
is strongly suppressed compared to ``sequential'' $3p$ emission via the
$^6$He($2^+$) state.

The channel potentials $\hat{V}_{ij}$ are obtained from $n$-$n$ and $\alpha$-$n$
potentials in the folding model using the realistic $^{6}$He WFs, e.g.\ from
\cite{Grigorenko:2020}. The ``SBB'' $\alpha$-$n$ interaction \cite{Sack:1954}
is used and simple effective $n$-$n$ interaction \cite{Cooper:1968}. The
diagonal potentials $\hat{V}_{11}$ for the $^{6}$He(g.s.)-$n$ channel and
$\hat{V}_{22}$ for the $^6$He($2^+$)-$n$ channel are used practically ``as is''
(some fine-tuning $\lesssim 5 \%$ can be applied). In contrast, the off-diagonal
potential $\hat{V}_{12}$ typically requires strong modification to provide
sufficient coupling to get realistic mixing values for the $^{7}$He g.s.; it
seem that this aspect of the $^{7}$He dynamics is beyond the folding
approximation.

An example of the coupled-channel studies is shown in Fig.\
\ref{fig:p-wave-off}. The obtained in this model population probabilities have
the following important features:

\begin{enumerate}

\item The experimental $^{7}$He $3/2^-_1$ g.s.\ width is naturally well
reproduced in this model.

\item The second $3/2^-_2$ state is found at relatively low energy ($E_T \approx
5.5$ MeV).

\item There is a strong deep in the population probability between the
$3/2^-_1$ and  $3/2^-_2$ states, which could be important for explanation of
correlations in $^{7}$He decay, see Sec.\ \ref{sec:p32-mod}.

\item It can be found in Fig.\ \ref{fig:p-wave-off} (b) that the off-shell
extrapolation, which was quite safe for the single-channel potential model, is
not working in the coupled-channel case.

\end{enumerate}


\section{Center-of-mass angular distributions}


The center-of-mass angular distributions provide the standard basis for 
spin-parity
identification in the direct reactions. For the $(d,p)$ reaction in inverse
kinematics the recoil protons escaping in the relatively broad ($5-30$ degrees)
angular range  in the backward direction in the lab system
correspond to small c.m.s.\ angular range ($2-11$ degrees).
Therefore, the use of the $(d,p)$ reaction in our studies has both
advantages and disadvantages.

The advantage for interpretation of our result is that population of states with
high $\Delta l$ should be suppressed in this experiment. For example the only
state, which is unambiguously identified in the excitation spectrum of $^{7}$He
is the $5/2^-$ state at $E \approx 3.4$ MeV, which decays with a large
probability via the $\isotope[6]{He}(2^+)$ channel
\cite{Korsheninnikov:1999,Wuosmaa:2008}. However, population of this state
requires $\Delta l=3$ compared to $\Delta l=1$ populating the $3/2^-$ and
$1/2^-$ states. So, it is quite natural to expect that only the $3/2^-$ and
$1/2^-$ resonant states of $^{7}$He are strongly populated in this experiment.

The disadvantage of the small angular range is that we are unable to make a
detailed comparison of the cross section angular distribution with reaction 
theory calculations to justify a specific spin-parity prescription. Nevertheless, 
we are going to demonstrate that
important conclusions are possible on the basis of the c.m.s.\ angular
distributions derived in this experiment.

\begin{figure}
\centering
\includegraphics[width=0.46\textwidth]{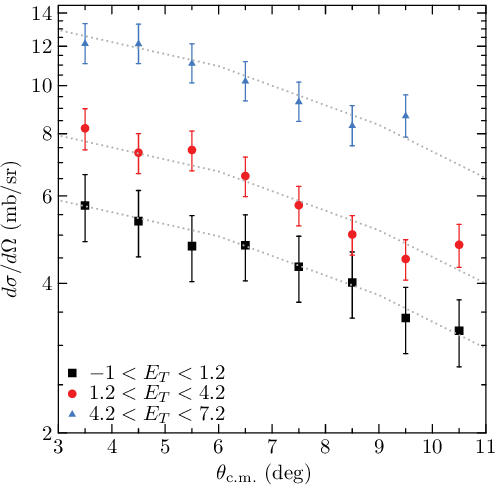}
\caption{\label{fig:angdis-comp-1}
  Differential cross section for the $^2$H($^{6}$He,$^1$H)$^{7}$He reaction
  for different ranges of the \isotope[7]{He} decay energy $E_T$.
  The dotted lines show the result of PWBA calculation for the \isotope[7]{He}
  $3/2^-$ g.s.\ (see also Fig.\ \ref{fig:angdis-comp-2}) scaled by the factors
  1, 1.35, and 2.2 to guide an eye.
}
\end{figure}

The experimental c.m.s.\ angular distributions for different excitation energy
ranges of $^{7}$He are shown in Fig.~\ref{fig:angdis-comp-1}. The data are
corrected for registration efficiency of the setup via the MC procedure. The 
obtained angular distributions can be considered as independent of
energy within the uncertainty of data, that may be a signal, that the same 
set of weights of different $\Delta l$ values is valid.

\begin{figure}
\centering
\includegraphics[width=0.45\textwidth]{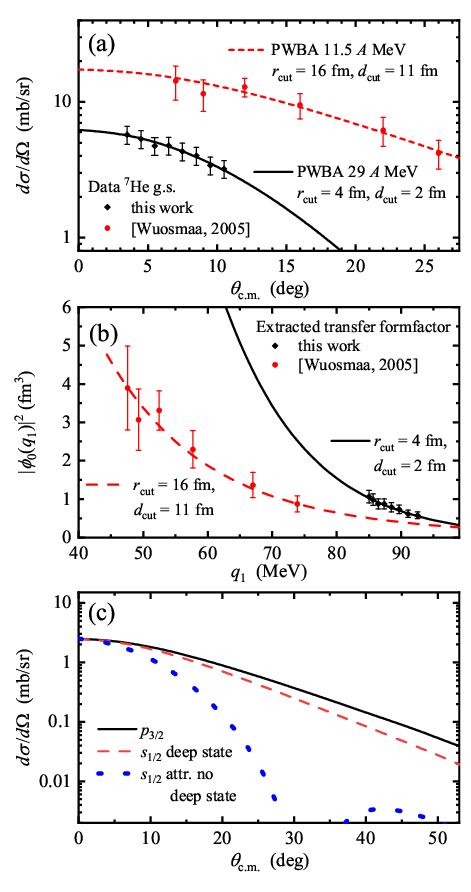}
\caption{(a) Comparison of the differential cross sections of the
$^2$H($^{6}$He,$^1$H)$^{7}$He reactions populating the \isotope[7]{He} g.s.\ as
obtained in this work and the work~\cite{Wuosmaa:2005}. Points show experimental
data. (b) Extracted squared transfer formfactor $|\phi_{0}(q_{1})|^2$ from both experimental 
data are shown by points. The curves correspond to the calculated deuteron formfactor
with the corresponding cut-off to show the best fit to the date.
(c) PWBA predictions for the angular distributions for  $\Delta l=0$ and
$\Delta l=1$ angular momentum transfers.}
\label{fig:angdis-comp-2}
\end{figure}

The comparison of experimental data with the PWBA calculations is given in Fig.\
\ref{fig:angdis-comp-2} (a). In its basic form, see Section \ref{sec:pwba},  the
PWBA model is not suitable for calculations of absolute cross sections. However,
there is a standard phenomenological modification of the model: the peripheral
Fermi-type cut-off for the ``projectile'' (deuteron) WF,
\[
\psi_0(r) \rightarrow
\frac{\psi_0(r)}{1-\exp[(r-r_{\text{cut}})/d_{\text{cut}}]} \,.
\]
It can be seen in Fig.~\ref{fig:angdis-comp-2} (a) that both the $^{7}$He  g.s.\
angular distributions of this work and of Ref.\ \cite{Wuosmaa:2005} can be reproduced
only by different parameter sets. Similary, in the DWBA studies of \cite{Bezbakh:2024}
somewhat different spectroscopic factor values $S=0.49$ and $S=0.39$ were
extracted for the 29 $A\, $MeV data and for the 11.5 $A\, $MeV data of Ref.\
\cite{Wuosmaa:2005}. Moreover it is demonstrated in Fig.~\ref{fig:angdis-comp-2} (b),
that no monotonic function $\phi_{0}$ exists, which can describe both data sets. 
So, the datasets can not be reconciled within typicaly
used direct reaction theoretical frameworks, and this is a
good indication, that absolute calibration of one of the data sets is actually wrong. This 
situation calls for a more precise experiment.

The stability of the angular distributions, demonstrated in
Fig.~\ref{fig:angdis-comp-1}, is quite natural if the obtained data is strongly
dominated by the same angular momentum transfer $\Delta l=1$ ($3/2^-$ g.s.,
$1/2^-$ and $3/2^-_2$ excited states), in the whole excitation energy range
$0<E_T<8$ MeV available in the experiment. The observation of the low-lying
$s_{1/2}$ resonance in $^{7}$He was declared in \cite{Golovkov:2024}, which
should also exhibit itself in the angular distributions. Unfortunately, it is
known, that specifically for the $(d,p)$ reactions the $\Delta l=0, 1, 2$ cross
sections have highly analogous profiles. It can be seen in
Fig.~\ref{fig:angdis-comp-2} (c) that to reliably distinguish the $\Delta l=0$
and $\Delta l=1$ contributions we should go to $\theta_{\text{c.m.}} \gtrsim
20^{\circ}$ and need a data with a high statistical confidence. So, one of the
ideas, how to prove or disprove the statement about the $s_{1/2}$ resonant state
in $^{7}$He is to perform a dedicated $(d,p)$ experiment, in which a broad
c.m.s.\ angular range is accessible.


\section{Correlation studies for the $^{7}$He spectrum}
\label{sec:corel}


It is shown in Section \ref{sec:pwba} that the momenta of \isotope[7]{He} decay
products should be strongly correlated with the \(\mathbf{q}_2\) direction.
In the experimental statistics there is a subset of events with
neutron coincidence and full kinematics. For the events in this subset the value
of \(\cos\theta_{k'}\) can be extracted from the experimental data and
\isotope[7]{He} decay energy measured with high resolution
($\lesssim 200$ keV FWHM), but this data has a very limited statistics and
a strong efficiency cut-off at $E_T \gtrsim 2$ MeV.

In contrast, the longitudinal velocity of the heavy decay fragment (and, 
correspondingly, the projection of \(\mathbf{k}'\) on the beam axis 
\(k'_{\parallel}\)) was measured for all events. For the events accepted by the 
setup of the experiment the equality
\[
  \cos(\theta_{k'}) \approx -k'_{\parallel}/k'\,,
\]
is precise within several percent. For this type of the data a broad 
excitation-energy range $E_T \lesssim 8$ MeV
is covered, but also the energy resolution is much lower ($\sim 600$ keV FWHM).
Due to low resolution of \(k'_{\parallel}\) and $E_T$ for this data the
correlations are available only in the form of backward-forward asymmetry in the
$^{7}$He frame.

Let us first discuss the latter type of the correlations.


\subsection{Backward-forward asymmetry in the distribution of core longitudinal
 momentum}
\label{sec:corel1}

\begin{figure}
\centering
\includegraphics{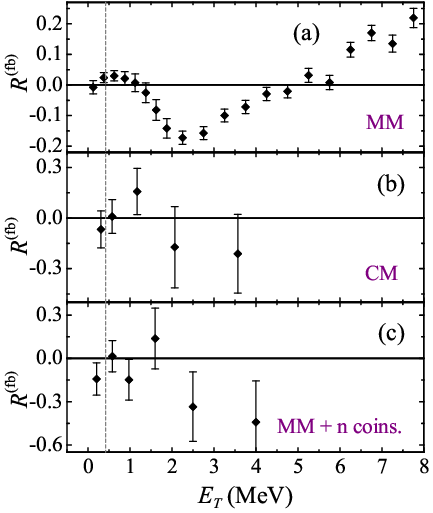}

\caption{\label{fig:ass-exp}
  Experimental backward-forward asymmetry for the $^{7}$He decay at different
  excitation energies for missing mass spectrum (a) and combined mass spectrum
  (b). The missing mass spectrum with additional neutron coincidence condition
  (c) is in reasonable agreement with the combined mass spectrum (b) indicating
  consistent calibrations of these two types of spectra. The vertical dashed
  line indicates the $^{7}$He g.s.\ position.
}

\end{figure}

In our discussion of correlations the ``forward'' direction is the direction of
\(\mathbf{q}_2\) and, correspondingly, \(\cos(\theta_{k'})>0\) or \(k'_{\parallel}<0\). The
experimental backward-forward asymmetry function is defined as
\[
R^{(\text{bf})}= \frac{W_{\text{f}}-W_{\text{b}}}{W_{\text{f}}+W_{\text{b}}}\,,
\]
see Fig.\ \ref{fig:ass-exp}. Important feature of this distribution is
rise from zero at $E_T\sim 0$ MeV, then quite abrupt change of sign at about
$E_T=1.5$ MeV, and then one more  change of sign at $E_T \sim 5$ MeV.

From Eq.\ (\ref{eq:corel-x-all}) the backward-forward asymmetry is obtained as
\begin{equation}
R^{(\text{bf})}= \frac{a_{s_{1/2}} \left[a_{p_{1/2}}
\cos(\delta_{s_{1/2}}^{p_{1/2}}) +
\sqrt{2} a_{p_{3/2}} \cos(\delta_{s_{1/2}}^{p_{3/2}}) \right] } {a^2_{s_{1/2}} +
a^2_{p_{1/2}} + a^2_{p_{3/2}}} \,.
\label{eq:rbf}
\end{equation}

\begin{figure}
\centering
\includegraphics[width=0.46\textwidth]{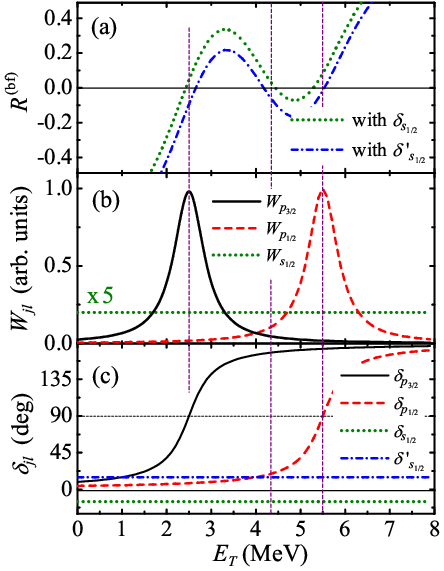}
\caption{Asymmetry for the simple model of $\{s_{1/2},p_{1/2},p_{3/2} \}$
interference: (a) asymmetry function; (b) population probabilities; (c) phase
shifts in the $^{6}$He-$n$ channel.}
\label{fig:ass-simple}
\end{figure}

Let us first demonstrate, how the observed backward-forward asymmetry can be
\emph{qualitatively} explained by a simple example of $\{s_{1/2},p_{1/2},p_{3/2}
\}$ correlation pattern for narrow $p$-wave states interference with flat
$s$-wave background, see Fig.\ \ref{fig:ass-simple}. The population
probabilities (\ref{eq:wjl}) and phase
shifts for this example are provided by the standard R-matrix parameterization
\[
\tan(\delta)= \frac{\Gamma}{2(E_r-E_T)} \,.
\]
We can find that the $R^{(\text{bf})}$ value is changing sign for the
\emph{first time} close to the $p_{3/2}$ resonance position (but not exactly at
resonance), at the point where
\begin{equation}
\delta_{s_{1/2}}^{p_{3/2}} \approx \pi/2 \,,
\label{eq:rbf-sign-1}
\end{equation}
since the $p_{1/2}$ contribution can be neglected here. The $R^{(\text{bf})}$
value is changing sign for the \emph{second time} around the point
\begin{equation}
W_{p_{1/2}} \approx 2 \, W_{p_{3/2}}\,,
\label{eq:rbf-sign-2}
\end{equation}
where the relative phase shift $\delta_{p_{3/2}} - \delta_{p_{1/2}} \approx
\pi$. The \emph{third time} the sign is changed in analogy with Eq.\
(\ref{eq:rbf-sign-1}) around the point
\begin{equation}
\delta_{s_{1/2}}^{p_{1/2}} \approx \pi/2 \,,
\label{eq:rbf-sign-3}
\end{equation}
where the $p_{3/2}$ contribution can be neglected.

We should also point that even a very small, visually negligible, contribution
of the $s$-wave configuration could be sufficient to produce the typical
observed asymmetry rate. The variation of the s-wave phase shift from attraction
to repulsion leads only to insignificant variation of the whole correlation
pattern, see $\delta_{s_{1/2}}$ and  $\delta'_{s_{1/2}}$ cases in Fig.\
\ref{fig:ass-simple}.

\begin{figure}
  \centering
\includegraphics[width=0.46\textwidth]{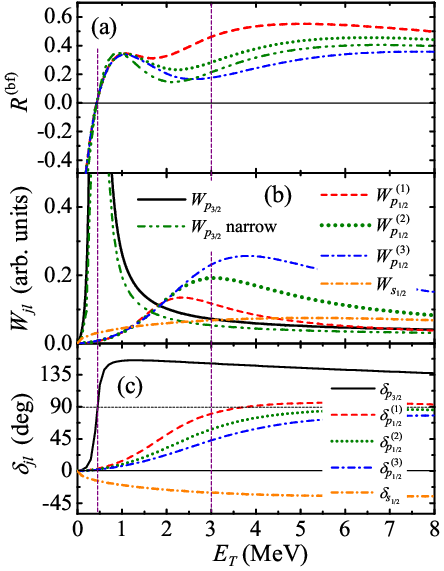}
\caption{Asymmetry for $\{s_{1/2},p_{1/2},p_{3/2} \}$ interference in PWBA model
with single-channel scattering states in the $^{6}$He-$n$ channel: (a) asymmetry
functions; (b) population probabilities; (c) phase shifts. The maximal variation
of the  $p_{1/2}$ state position admissible by the data of Fig.\
\ref{fig:comp-spec-all} is performed. }
\label{fig:ass-poten}
\end{figure}

The use of realistic profiles provided by potential model for the $^{7}$He
$T$-matrix gives qualitatively similar behavior, but do not provide quantitative
agreement, see Fig.\ \ref{fig:ass-poten}. There is a kink in the asymmetry
function in between the $p_{3/2}$ and $p_{1/2}$ resonances, but the value is not
falling towards zero: the effect of the $p_{1/2}$ state seem to be too small and
it should be enlarged to achieve agreement with experiment. However, it can be
found that contributions of the $p_{1/2}$ resonance are already on the upper
limit admissible by the experimental MM spectrum, see Fig.\
\ref{fig:comp-spec-all}. Alternatively we may inquire, whether the contribution
of the $p_{3/2}$ state may be reduced at $E_T>1-1.5$ MeV. It can be seen in
Fig.\ \ref{fig:ass-poten} (green double-dotted curve) that within the potential
approach even the unrealistically strong variation of the $p_{3/2}$ g.s.\
properties does not lead to sufficiently strong variation of the high-energy
tail of this state.

We have found three possible explanations for the data, as listed in the
following Sections.


\subsubsection{Phase shift variation}
\label{sec:delta-mod}

\begin{figure}
\centering
\includegraphics[width=0.46\textwidth]{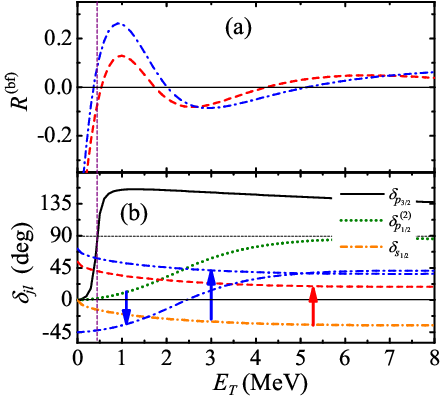}
\caption{Asymmetry for $\{s_{1/2},p_{1/2},p_{3/2} \}$ interference in PWBA model
with modified phase shift convention: (a) asymmetry functions; (b) two variants
of of phase shifts modification of Eq.\ (\ref{eq:phase-var}) are shown by red
and blue curves and correspondingly colored arrows.}
\label{fig:ass-delta-mod}
\end{figure}

The phase behavior of the $T$-matrix elements is provided in PWBA as elastic
phase shift in the $^{7}$He channel, see Eq.\ (\ref{eq:amp-repres}). However, as
we have already mentioned in the Section \ref{sec:pwba}, that this is not
necessarily true in situation with some general reaction mechanism. It can be
found that modifications of the T-matrix phase behavior
\begin{eqnarray}
\delta_{s_{1/2}} & \rightarrow & \delta_{s_{1/2}} + 0.3 \pi \, , \nonumber \\
\{\delta_{s_{1/2}},\delta_{p_{1/2}} \} & \rightarrow & \{\delta_{s_{1/2}} + 0.4
\pi \,,\; \delta_{p_{1/2}} - 0.25 \pi \} \, , \qquad
\label{eq:phase-var}
\end{eqnarray}
lead to results reasonably consistent with experimental observations, see
Fig.\ \ref{fig:ass-delta-mod}. The first variant of
modification in Eq.\ (\ref{eq:phase-var}) looks more realistic, since the
relative phase shift of $s$-wave and $p$-wave $T$-matrices looks quite
reasonable, say within DWBA, while the sizable relative phase shift of the
$p_{3/2}$ and $p_{1/2}$ components is more complicated to justify.

The considered phase shift modification is beyond the simplistic PWBA discussed
in this work. The question can be asked how realistic is such a modification,
which we leave here as an opened question to reaction theory.


\subsubsection{Suppression of $p_{3/2}$ above $E_T=1$ MeV}
\label{sec:p32-mod}

\begin{figure}
\centering
\includegraphics[width=0.46\textwidth]{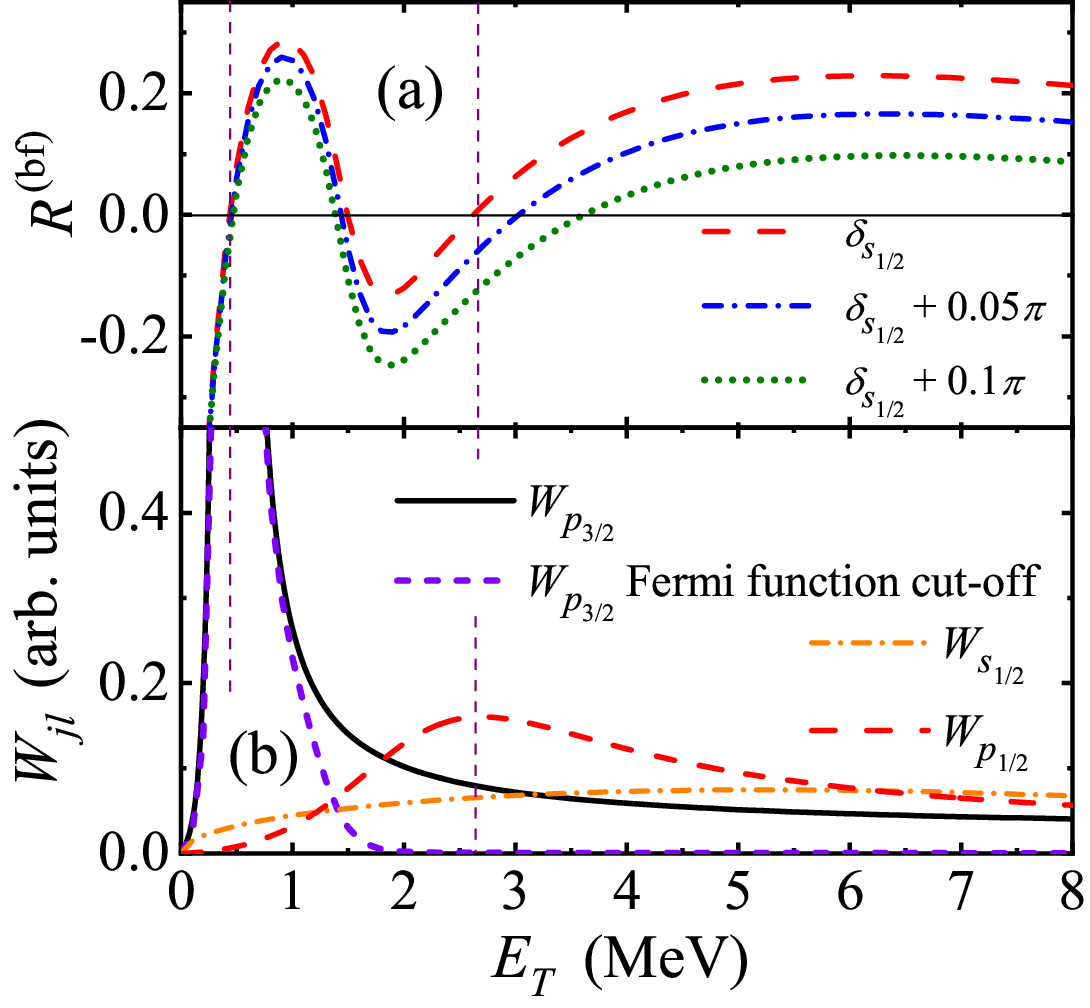}
\caption{Asymmetry for $\{s_{1/2},p_{1/2},p_{3/2} \}$ interference in PWBA model
with single-channel scattering states in the $^{6}$He-$n$ channel, but with
modified (``Fermi function cut-off'') high-energy tail of the $p_{3/2}$ resonance:
(a) asymmetry function; (b) population probabilities. Small modifications of the
$s_{1/2}$ phase shifts may lead to improved agreement with the data.}
\label{fig:ass-p32-mod}
\end{figure}

Another possible explanation of the experimental asymmetry function, which
naturally stems from Eq.\ (\ref{eq:rbf}) is connected with modification of the
high-energy tail of the $p_{3/2}$ resonance. To check such possibility we cut off
the high energy $p_{3/2}$ tail via multiplication the $W_{3/2,1}$ 
population probabilities by Fermi function with
energy parameter 1.3 MeV and width parameter 0.15 MeV. The results of
calculations are shown in Fig.\ \ref{fig:ass-p32-mod}. There is good qualitative
agreement at once and also a quantitative agreement with the data may be
achieved by small variations of the phase shifts
\begin{eqnarray}
\delta_{s_{1/2}} \rightarrow \delta_{s_{1/2}} + 0.05 \pi \,, \nonumber
\\
\delta_{s_{1/2}} \rightarrow \delta_{s_{1/2}} + 0.1 \pi \,.
\label{eq:phase-var-2}
\end{eqnarray}
%

Such a suppression of the high-energy tail of the $p_{3/2}$ resonance may find a
natural explanation as a threshold effect in the $^{7}$He continuum. Indeed, the
asymmetry function $R^{(\text{bf})}$ is changing sign at around $E_T=1.5$ MeV,
which is suspiciously close to the $^{6}$He($2^+$)+$n$ channel threshold at
$E_T=1.8$ MeV. To take this effect into account we need a coupled-channel
calculations. Example of such calculations is shown in Fig.\
\ref{fig:p-wave-off}, where a strong suppression of the $^{6}$He($0^+$)+$n$
channel population is found around the $^{6}$He($2^+$)+$n$ threshold energy. We
find this possibility the best candidate for reasonable explanation of
experimental observations.

%
%


\subsubsection{The $s_{1/2}$ resonant state in $^{7}$He}
\label{sec:s12-mod}

\begin{figure}
\centering
\includegraphics[width=0.46\textwidth]{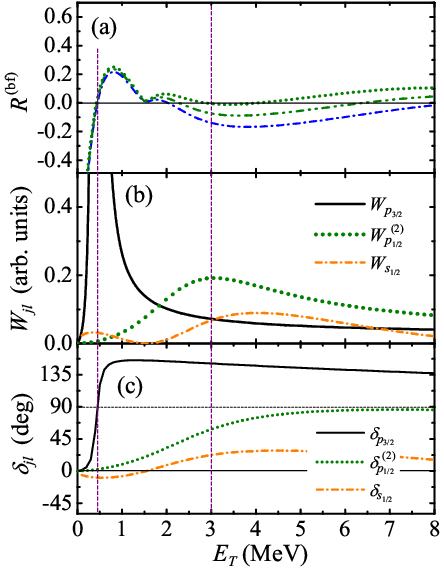}
\caption{Asymmetry for $\{s_{1/2},p_{1/2},p_{3/2} \}$ interference in PWBA model
with single-channel scattering states in the $p$-wave channels, but with
resonant contribution in the $s_{1/2}$ channel: (a) asymmetry function; (b)
population probabilities; (c) phase shifts.}
\label{fig:ass-s12-mod}
\end{figure}

Another possible explanation of the experimental asymmetry function, which
naturally stems from Eq.\ (\ref{eq:rbf}) is modification of the $s_{1/2}$
continuum component, which can be connected with  $s_{1/2}$ resonance somewhere
in $^{7}$He continuum. The data interpretation in these terms was proposed in
paper \cite{Golovkov:2024}. For this interpretation the R-matrix
parameterization was used providing the phase shift
\begin{equation}
\delta_{s_{1/2}}(E_T) = \arctan \left[\frac{S_{s_{1/2}}\,k'}{4 M_{tp}r_{ch}(E_r
- E_T)} \right] - k' r_{ch} \,.
\label{eq:rmatr-s-wave-res}
\end{equation}
The first term here is R-matrix resonance phase shift parameterization for the
$s$-wave and the second term is just a hard-sphere-scattering phase shift. The
calculation results with $E_r=2.0$ MeV, $r_{ch}=4.5$ fm, $S_{s_{1/2}}=1.3$ and
with
$s$-wave T-matrix calculated on-shell are shown in Fig.\ \ref{fig:ass-s12-mod}.
The resonance width which can be formally associated with such parameterization
is $\Gamma =S_{s_{1/2}} k'r_{ch} = 2$ MeV. Qualitatively reasonable description
of the observed asymmetry can be achieved.

This interpretation is based on two assumptions: (i) The off-shell interpolation
of the elastic cross section induced by Eq.\
(\ref{eq:rmatr-s-wave-res}) was assumed to be trivial: the Authors of
\cite{Golovkov:2024} followed Ref.\ \cite{Fuchs:1972} where the $s$-wave
off-shell interpolation is just constant; (ii) the parameters in the R-matrix
expression Eq.\ (\ref{eq:rmatr-s-wave-res}) can be interpreted as actual  
resonance properties of the $^{7}$He continuum.

\begin{figure}
\centering
\includegraphics[width=0.47\textwidth]{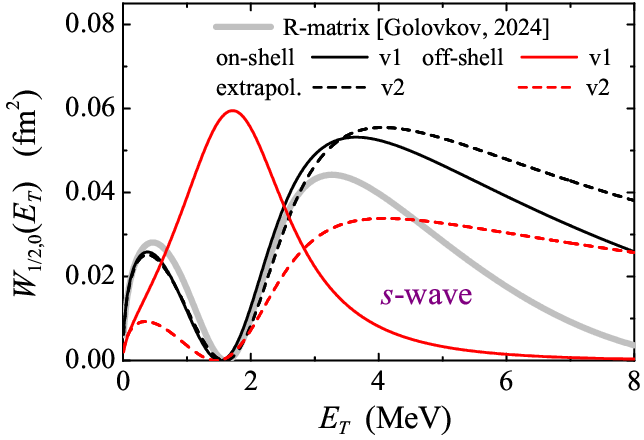}
\caption{The population probabilities Eq.\ (\ref{eq:wjl}) presuming existence of
a resonant $s$-wave state are calculated off-shell at
$\theta_{\text{c.m.}}=8^{\circ}$, and extrapolated from the on-shell value by
Eq.\ (\ref{eq:w-off1}). The R-matrix-parameterized value used in
\cite{Golovkov:2024} is shown by thick solid gray curve. Black solid and dashed curves
shows two versions of the coupled-channel calculations practically coinciding
with the latter on-shell at $E_T \lesssim 3$ MeV.}
\label{fig:s-wave-off-cc}
\end{figure}

Concerning these assumptions we can comment the follows:

\begin{enumerate}

\item The off-shell interpolation of the $s$-wave T-matrix is not constant even
in the simple potential model. It was shown by direct calculation in Fig.\
\ref{fig:s-wave-off} that the off-shell modification of the T-matrix amplitudes
can be as large as factor $2-3$ both up and down depending on particular
properties of the $^{7}$He $s$-wave interaction.

\item A resonance in $s$-wave (in contrast with virtual state) can be obtained
only in the coupled-channel formulation of the scattering problem. For such
formulation the $s$-wave pole can arise due to coupling with resonant state in
some, otherwise closed, channel. We performed the phenomenological studies
within  coupled-channel model trying to find the situation which can lead to
phase shift behavior analogous to the one shown in Fig.\ \ref{fig:ass-s12-mod}.
It was possible, but the deduced resonance properties of the continuum were 
found to be very different: $E_r \sim 6.5$ MeV and $\Gamma \sim 6$ MeV. This is 
the issue which
is very familiar to practitioners of the R-matrix studies: for broad states the
R-matrix parameters can be not easy to relate to physical observables. The
$s$-wave continuum just appeared especially weird in this sense.

\item In the coupled-channel model the off-shell behavior of the T-matrix can be
straightforwardly obtained. The results of such calculations are shown in Fig.\
\ref{fig:s-wave-off-cc}. One may see that the off-shell behavior in such a model
is complicated and \emph{qualitatively different} from the on-shell behavior.
So, if we really deal with the resonant state in $s$-wave, the T-matrix should
be only computed: there is no way to infer it from the on-shell (elastic)
scattering T-matrix.

\end{enumerate}

Thus, the interpretation proposed in \cite{Golovkov:2024} is ``phenomenologically
possible'' as off-shell T-matrix parameterization describing the data, but this interpretation 
can not be unambiguously related to ``real physics'' of the $^{7}$He continuum
described by the on-shell T-matrix.

\subsection{Correlations around the $^{7}$He g.s.}
\label{subsec:corel2}

\begin{figure}
\centering
\includegraphics{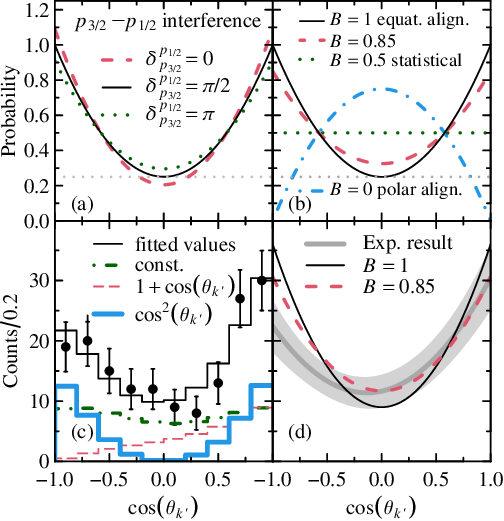}
\caption{\label{fig:he7-gs-corel}
  Angular distributions for the $^{7}$He g.s. decay, see Eq.\ (\ref{eq:corel-x-all}).
  (a) Theoretical angular distribution for the case of $p_{3/2}$-$p_{1/2}$ 
  mixing and complete equatorial alignment $B=1$ for the $p_{3/2}$ amplitude; 
  average relative   weight of $p_{1/2}$ is around $10^{-3}$.
  (b) Distributions for pure $p_{3/2}$ state with different alignment $B$ 
  values.
  (c) Experimental data are shown by circles. The solid histogram shows the 
  result of fit based on Eq.~\eqref{eq:parabol};
  other histograms show contributions of individual terms of \eqref{eq:parabol}.
  (d) Comparison of the distribution obtained from the regression analysis
  with results of panel (b) shows that a very high alignment is required with 
  confident limit $B>0.85$.}
\end{figure}

For this energy range the angular distribution for the $^{6}$He-$n$ relative
motion can be reconstructed. We make it in the $^{7}$He c.m.\ frame with
$Z||\mathbf{q}_2$, where PWBA promise the most expressed correlation pattern.
The data integrated over the $E_T=0.2-0.6$ MeV range is shown in
Fig.~\ref{fig:he7-gs-corel} (c). For the $^{7}$He g.s.\ the description of the
angular distribution can be confidently attributed to the
$\{s_{1/2},p_{1/2},p_{3/2} \}$ interference. Then the whole distribution should
be general parabolic shape
\begin{equation}
\frac{dW}{dx}=c_0 \frac{1}{2} + c_1 \frac{x+1}{2} + c_2 \frac{3x^2}{2}\,,
\quad x = \cos(\theta_{k'})
\label{eq:parabol}
\end{equation}
The distortions of individual terms from Eq.\ (\ref{eq:parabol}) by the
experimental setup were studied by using the MC procedure and we may see in
Fig.~\ref{fig:he7-gs-corel} (c) that these distortions are modest.

For the pure $p_{3/2}$ the PWBA model Eq.\ (\ref{eq:corel-x-all}) predicts the
expressed parabolic profile with ``hill-to-valley'' ratio equal 4, see
Fig.~\ref{fig:he7-gs-corel} (a,b,d) black solid curve. Let's discuss meaning of
this fact and whether some additional information could be extracted from this
distribution.

\begin{figure}
\centering
\includegraphics[width=\linewidth]{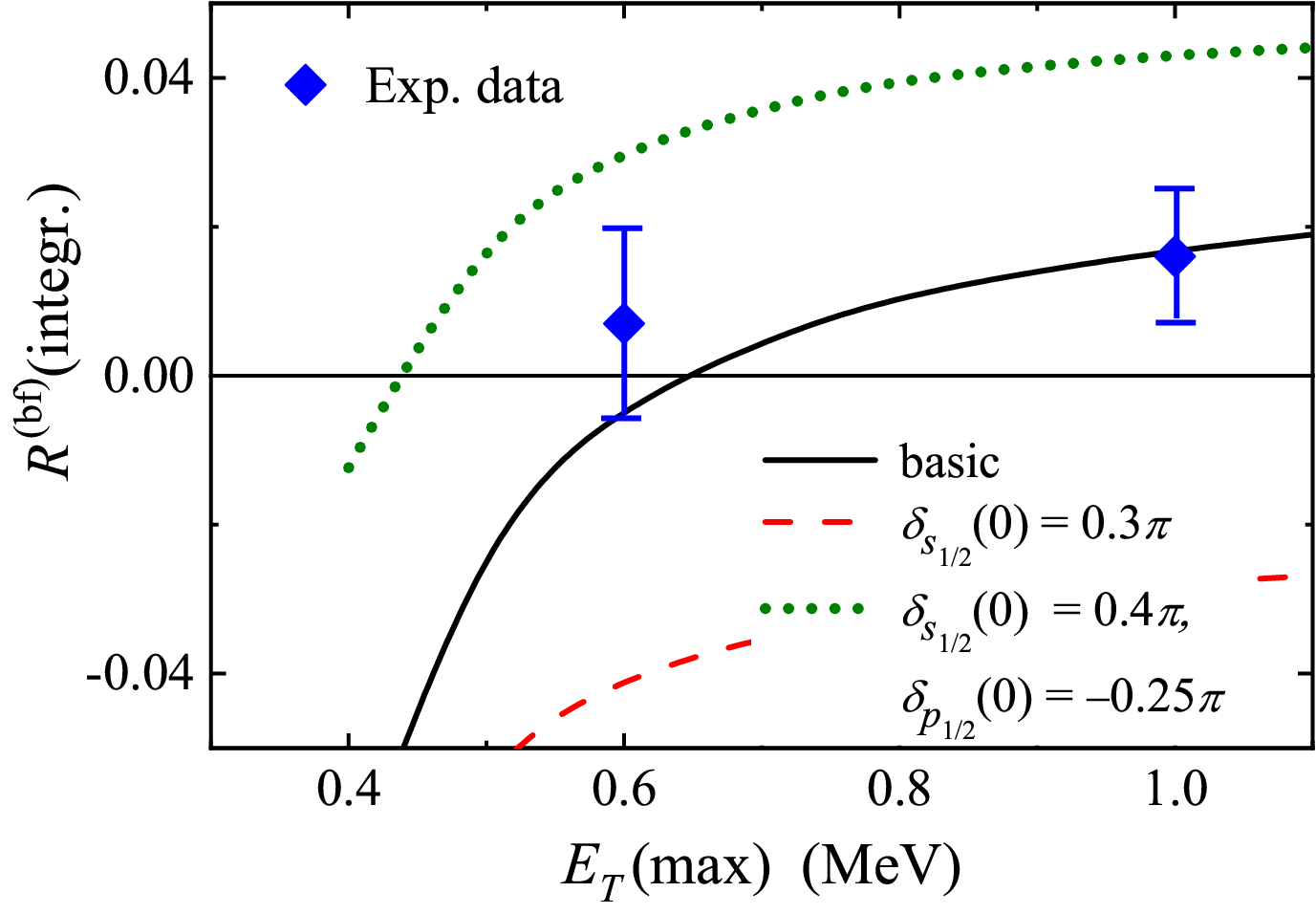}
\caption{The asymmetry function integrated in the range $E_T=\{0.2,E_T(\max)\}$.
The ``basic'' case is obtained from green dotted curve from Fig.\
\ref{fig:ass-p32-mod}. The other two cases are obtained by phase modification as
in Eq.\ (\ref{eq:phase-var}). }
\label{fig:ass-integ}
\end{figure}

\begin{figure}
\centering
\includegraphics[width=\linewidth]{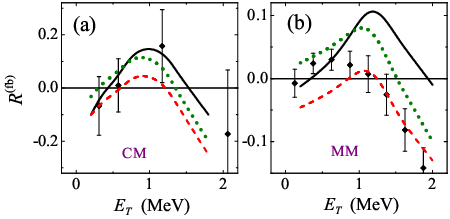}
\caption{The asymmetry function taking into account energy resolution is
compared to the data. For the c.m.\ spectrum (a) the resolution is $\sim 0.2$
MeV FWHM, and MM spectrum (b) the resolution is $\sim 0.62$ MeV. The curves are
the same as in Fig.\ \ref{fig:ass-integ}.}
\label{fig:ass-low-comp}
\end{figure}

Some asymmetry can be found in the distribution Fig.~\ref{fig:he7-gs-corel} (c).
However, this asymmetry is actually difficult to relate to physics of the 
experiment.
The integrated asymmetry effect connected with $p_{3/2}$-$s_{1/2}$ interference
at the $^{7}$He $p_{3/2}$ g.s.\ is quite small because asymmetry is changing
sign around the $p_{3/2}$ resonant energy and the observed small value is result
of fine compensation of large values from below and from above of the resonance.
Fig.\ \ref{fig:ass-integ} shows how this effect depends on the integration
range. It is clear that in any experiment, where the energy resolution is lower
than the width of the $^{7}$He g.s.\ such an asymmetry actually shows the
integration energy range which we can afford because of resolution and
statistics. It can be also found from Fig.\ \ref{fig:ass-integ} that there is
large sensitivity of the integrated asymmetry value on the phase convention.
Potentially, this property can be used to determine the phase convention
experimentally. At the moment the quality of the data is not sufficient (the
error bars shown in this plot are pure statistical, the actual uncertainty is
larger). The dependence of the measured asymmetry on the experimental resolution
is further elaborated in Fig.\ \ref{fig:ass-low-comp}. Again we can find that
potentially, the asymmetry function can be used to determine the phase
convention experimentally. However, at the moment the comparison of  Fig.\
\ref{fig:ass-low-comp} can lead only to qualitative conclusions as the c.m.\
spectrum has good enough resolution, but low-statistics data with large error
bars, while the MM spectrum reliable statistically suffer from large effect of
energy resolution.

Another aspect of the angular distributions in Fig.~\ref{fig:he7-gs-corel} is
connected with $p_{3/2}$-$p_{1/2}$ interference and $p_{3/2}$ alignment
properties.

The $p_{3/2}$-$p_{1/2}$ interference averaged over the energy integration range
is illustrated in Fig.~\ref{fig:he7-gs-corel} (a). The effect is small, whatever
relative $p_{3/2}$-$p_{1/2}$ phase is assumed. Actually we expect this phase
close to $\pi/2$ at the resonance position, and therefore small
interference effects anyhow. So, here no sensitivity is found and no information
can be extracted.

Situation with alignment is more interesting. Among configurations $p_{3/2}$,
$p_{1/2}$, and $s_{1/2}$ only the $p_{3/2}$ one may be spin-aligned.
It is convenient to discuss the $p_{3/2}$ alignment in terms of the alignment
coefficient
\[
B = \frac{W_{M=\pm 1/2}}{W_{M=\pm 1/2}+ W_{M=\pm 3/2}}  \,.
\]
It is clear that for pure $p_{3/2}$ configuration, the $B= 0.5$ alignment lead
to the isotropic angular distribution (this is case of ``statistical''
population of magnetic substates), see Fig.\ \ref{fig:he7-gs-corel} (b), green
dotted curve. The complete ``equatorial'' alignment with $B= 1$ leads to an
expressed parabolic concave profile with ``baseline'' equal to 0.25, see Fig.\
\ref{fig:he7-gs-corel} (b), solid black curve.  The
``polar'' alignment with $B \sim 0$ leads to parabolic convex profiles.
The latter profile is clearly in a strong disagreement with the data. The
experimental data Fig.\ \ref{fig:he7-gs-corel} (d) clearly requires pure or
practically pure equatorial alignment with $B > 0.85$. Pure equatorial alignment
is trivially
predicted in PWBA or similar models, which can be qualitatively attributed to
the  single-step (or ``single-pole'') reaction mechanism. So, the observed
angular distribution confirming the near-perfect equatorial alignment can be
seen as strong support of the applicability of the used theoretical approach.

So, to finalize this point, the observed angular correlations for the $^{7}$He
g.s.\ decay provide a strong confirmation for a
``single-pole'' single-step reaction mechanism leading to population of $^{7}$He
ground state. We have shown that the data of this kind, but higher quality can
be used to establish solidly the phase convention for $s_{1/2}$-$p_{3/2}$
interference in proximity of $^{7}$He ground state.


\section{Outlook}


The results obtained in this work show nice prospects of the discussed
correlation method for detailed and precise studies of the $^{7}$He continuum.
The following issues may be addressed in the forthcoming studies of the system.

\begin{figure}
\centering
\includegraphics[width=0.48\textwidth]{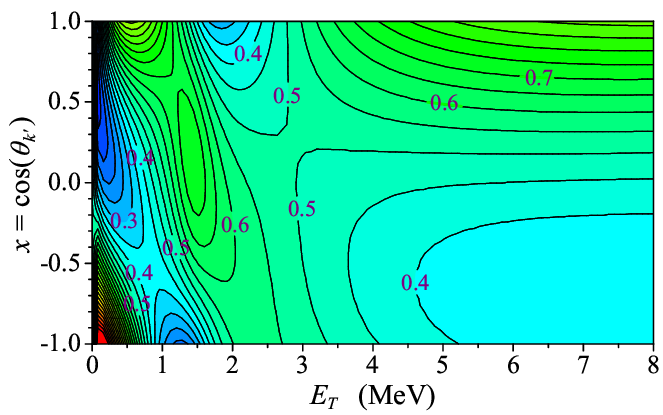}
\caption{Complete correlation function $\frac{d^2 \sigma}{dE_Tdx}/\frac{d \sigma}{dE_T}$ 
normalized to unit at each $E_T$ value for
$^{6}$He-$n$ decay. The shown case corresponds to phase prescription of Fig. \ref{fig:ass-p32-mod} (green dotted curve),
which qualitatively best fits the experimental data shown in Fig. \ref{fig:ass-exp}(a).}

\label{fig:comp-cor-p32-mod}
\end{figure}

\begin{enumerate}

\item The use of the ``combined mass'' approach allows to drastically
improve the energy resolution of the $^{7}$He spectrum. Aiming experiment with
high statistics we may pretend to derive the $^{7}$He g.s.\ width with precision
$\Delta \Gamma \sim 10-15$ keV. The $^{7}$He g.s.\ is well known to be not a
single particle state with the dominant $^{6}\text{He}(2^+)$+$n$ configuration
with spectroscopic factors varying in different studies in the range $0.35-0.65$
(e.g.\ \cite{Renzi:2016,Fortune:2018}). The improved width data would allow to
elaborate this question.

\item The variants of explanation for the asymmetry function formulated
in Sections \ref{sec:delta-mod}, \ref{sec:p32-mod}, and \ref{sec:s12-mod} are to
certain extent connected with the limited character of information provided by
asymmetry. This drawback could be overcome in a more sophisticated experiment,
where high-quality complete correlation information will be available in a broad
energy range. The illustration of the predicted complete correlation pattern is
given in Fig.\ \ref{fig:comp-cor-p32-mod}. The important aspect of
information which is missing in the current data is transition convex-concave
for the parabolic component of the $^{6}$He-$n$ angular distribution. The
corresponding part of the cross section is provided in PWBA as
\begin{equation}
\frac{d^2 \sigma^{(2)}}{dE_Tdx} \propto \, x^2 \,a_{p_{3/2}} \, \left[ 
a_{p_{3/2}} +  2 \sqrt{2} \, a_{p_{1/2}} \,\cos(\delta_{p_{1/2}}^{p_{3/2}})
\right]  \,.
\label{eq:corel-x-power-2}
\end{equation}
and it is clear that the coefficient at the $x^2$ term is changing sign in the
points providing different information compared to the asymmetry coefficient
Eq.\ (\ref{eq:rbf}).

\item  The limitations on the properties of the $1/2^-$ resonance are
obtained in this work are not very restrictive. There are two main reasons.
(i) Unresolved contribution of the $^{6}\text{He}(2^+)$-$n$ inelastic
channel in the experiment. This issue can be easily resolved in a more
``accurate'' design of the experimental setup providing the clear identification
of $^{4}$He and $^{6}$He. (ii) Uncertainty in the population of the $s$-wave
continuum.  This issue is more complicated as main qualitative properties of the
correlations retain, for example, for 10-fold (even 50-fold) reduction of the
$s$-wave continuum population compared to the practically maximal possible
contribution used in the current analysis. However, potentially it can also be
resolved if the precise data of the type shown in Fig.\
\ref{fig:comp-cor-p32-mod} is available.

\end{enumerate}


\section{Conclusions}


The $^{7}$He continuum states were studied in the
$^{2}$H($^{6}$He,$^{1}$H)$^{7}$He reaction at 29 $A\,$MeV. As compared to
previous studies in the same reaction \cite{Golovkov:2001,Wuosmaa:2005} this
experiment provides higher statistics, higher resolution, and larger excitation
energy coverage. The details of the analysis procedure are somewhat different
from those published in \cite{Bezbakh:2024,Golovkov:2024} and there are
important differences in interpretation of the data. The interpretation of the
data in this work is based on extensive PWBA plus coupled-channel studies.  They
are indicating extreme importance of the careful treatment of the off-shell
effects for understanding of the correlation patterns induced in $^{7}$He by the
$(d,p)$ reaction.

The population cross section for the $^{7}$He g.s.\ is found to be inconsistent 
with the data \cite{Wuosmaa:2005} obtained at 11.5 $A\,$MeV. The c.m.\ angular
distributions for the reaction are practically independent on the $^{7}$He 
excitation energy, supporting the same reaction mechanism and the same dominant 
$\Delta l=1$ up to $E_T=8$ MeV.

The $^{7}$He $3/2^-$ g.s.\ properties are established as $E_r = 0.41(2)$ MeV and 
$\Gamma = 0.14(5)$ keV. Angular distribution for the $^{6}$He-$n$ decay of
this state can be explained by a strong spin alignment induced by a reaction
mechanism. The equatorial spin alignment is a natural feature of the single-step
single-pole direct reaction models including PWBA. Thus this fact is a solid
confirmation of the robustness of our theoretical considerations.

The correlation information for the  higher-lying $^{7}$He excitations is 
available as backward-forward asymmetry for the $^{6}$He-$n$ decay in the 
$^{7}$He frame aligned with transferred momentum $\mathbf{q}_2$. The asymmetry 
function has an expressed profile with three sign-changing points (the first of 
them is very likely from the data and two others are reliably observed). Such a 
behavior of this function may be explained by using quite restrictive 
assumptions and only three such explanations were found:

\begin{enumerate}

\item Strong phase variation (compared to PWBA) due to more complicated reaction
mechanism. Possibility of such an interpretation could be confirmed or dismissed
by the further advanced reaction theory studies.

\item Suppression of $3/2^-$ continuum above $E_T=1$ MeV. This interpretation
gets support from the coupled channel calculations as a threshold effect in
proximity to the $^{6}$He($2^+$)+$n$ channel threshold.

\item The existence of the $s_{1/2}$ resonant state in $^{7}$He was declared in
Ref.\ \cite{Golovkov:2024} with $E_r \approx 2.0 $, $\Gamma \approx 2.0$ MeV.
This interpretation is in contradiction with the coupled-channel calculations,
which predict extremely strong off-shell effects in this case, providing
qualitatively different picture after off-shell correction. No special need for
the $s_{1/2}$ resonant state is found in the data analysis of this work.

\end{enumerate}

In all the above scenarios the first excited state of $^{7}$He is (quite
naturally) $1/2^-$ state with peak positions confined to a range $E_r=2.2-3.0$
MeV with ``preferred'' value around 2.6 MeV. There is indication on the second
$3/2^-$ state in the data with $E_r \sim 4.5$ MeV and the lower resonance energy
limit $E_r \gtrsim 3.5$ MeV.

In this work it was demonstrated that the discussed correlation method has the potential to
solve the listed  burning questions of the $^{7}$He continuum. However it is clear, that prospective 
experiments should have higher statistics, higher energy resolution and unambiguous 
identification of decay channels.


\begin{acknowledgments}
We are grateful to Prof. A.S.\ Fomichev for important discussions. The
research was supported in part in the framework of scientific program of the
Russian National Center for Physics and Mathematics, topic number 6 ``Nuclear
and radiation physics'' (2023--2025 stage). We acknowledge the interest and
support of this activity from Profs. Yu.Ts.\ Oganessian, S.N.\ Dmitriev and
S.I.\ Sidorchuk.
\end{acknowledgments}


\bibliographystyle{apsrev4-2}
\bibliography{corr-studies-2.bbl}
\end{document}